\documentclass{aastex}          
\usepackage{spr-astr-addons}    

\usepackage{hyperref}
\usepackage{graphicx}
\usepackage{amsmath,amssymb}
\usepackage{bm} 
\usepackage{widetext}
\usepackage{multirow}

\newcommand{\avg}[1]{\langle #1 \rangle}
\newcommand{\LRavg}[1]{\left\langle #1 \right\rangle}

\newcommand{\dd}[0]{\ensuremath{\mathrm{d}}}%
\newcommand{\ee}[0]{\ensuremath{\mathrm{e}}}%
\newcommand*{\ii}{\imath}

\renewcommand{\vec}[1]{\ensuremath{\bm{#1}}}%
\newcommand{\vhat}[1]{\ensuremath{\hat{\bm{#1}}}}%
\newcommand{\vA}[0]{\ensuremath{v_{\text{A}}}}%
\newcommand{\Breg}[0]{\ensuremath{\langle \vec{B} \rangle}}%
\newcommand{\Bregular}[0]{\ensuremath{B_z}}%
\newcommand{\kparr}[0]{\kappa_{\parallel}}
\newcommand{\kperp}[0]{\kappa_{\perp}}

\usepackage{pifont} 
\newcommand{\cmark}{\ding{51}}%
\newcommand{\xmark}{\ding{55}}%

\usepackage[pscoord]{eso-pic}
\newcommand{\placetextbox}[3]{
	\setbox0=\hbox{#3}
	\AddToShipoutPictureFG*{
		\put(\LenToUnit{#1\paperwidth},\LenToUnit{#2\paperheight}){\vtop{{\null}\makebox[0pt][c]{#3}}}
	}
}

\begin{document}

\title{Test particle simulations of cosmic rays}

\shorttitle{Test particle simulations}
\shortauthors{Philipp Mertsch}

\author{Philipp Mertsch} 
\affil{Institute for Theoretical Physics and Cosmology (TTK), RWTH Aachen University, Sommerfeldstr.\ 16, 52074 Aachen, Germany}
\email{pmertsch@physik.rwth-aachen.de} 

\begin{abstract}
Modelling of cosmic ray transport and interpretation of cosmic ray data ultimately rely on a solid understanding of the interactions of charged particles with turbulent magnetic fields. The paradigm over the last 50 years has been the so-called quasi-linear theory, despite some well-known issues. In the absence of a widely accepted extension of quasi-linear theory, wave-particle interactions must also be studied in numerical simulations where the equations of motion are directly solved in a realisation of the turbulent magnetic field. The applications of such test particle simulations of cosmic rays are manifold: testing transport theories, computing parameters like diffusion coefficients or making predictions for phenomena beyond standard diffusion theories, e.g.\ for cosmic ray small-scale anisotropies. In this \textit{review}, we seek to give a low-level introduction to test particle simulations of cosmic rays, enabling readers to perform their own test particle simulations. We start with a review of quasi-linear theory, highlighting some of its issues and suggested extensions. Next, we summarise the state-of-the-art in test particle simulations and give concrete recipes for generating synthetic turbulence. We present a couple of examples for applications of such simulations and comment on an important conceptual detail in the backtracking of particles.
\end{abstract}

\keywords{Cosmic ray theory, cosmic ray transport, wave-particle interactions, magnetic turbulence, computer simulations}

\placetextbox{0.967}{0.86}{\normalsize \normalfont TTK-19-37}

\section{Introduction}

Cosmic rays (CRs), that is the population of charged, relativistic particles with non-thermal spectra, are ubiquitous in the Universe. They pervade systems of all sizes, from stellar systems to whole galaxies, from galaxy clusters to the intercluster medium. See~\citet{Ginzburg:1990sk,Strong:2007nh,2015ARA&A..53..199G,2011ARA&A..49..119K} for reviews on Galactic and extra-galactic cosmic rays. CRs are not only responsible for genuinely non-thermal phenomena: the fluxes of CRs observed at Earth, the non-thermal emission of radio, X-ray and gamma-ray sources or the diffuse Galactic and extragalactic emission; but CRs oftentimes have energy densities comparable or even superior to other components, like the thermal gas, magnetic fields or radiation backgrounds. As such, CRs can contribute to the pressure equilibrium or even drive large-scale outflows, e.g.~\citealt{Everett:2007dw,2013ApJ...777L..38H,2016ApJ...827L..29S,2016MNRAS.462.4227R}. At the largest scales, it has been suggested that CRs (or gamma-rays from blazars) contribute to the heating of the Universe at redshifts as high as $z \sim 10$~\citep{1993MNRAS.265..241N,2015MNRAS.454.3464S,2017MNRAS.469..416L}.

Any detailed modelling of CRs relies on understanding transport in coordinate and momentum space. For instance, modelling the locally observed CRs involves their propagation from the sources to the observer. It is believed that diffusion is the dominant process in shaping the spectra, both during shock or stochastic acceleration inside the sources and during their transport from the sources. Indeed for Galactic CRs the most important effect, that is the softening of the observed spectra with respect to the source spectra and the relative softness of so-called secondary species (e.g.\ boron) with respect to so-called primary species (e.g.\ carbon), can be explained with a rigidity-dependent diffusion coefficient. (See~\citet{Gabici:2019jvz} for a recent review of the challenges to this picture.) Scrutinizing this picture and improving upon it requires a better, more refined understanding of spatial transport. A prominent example is the issue of small-scale anisotropies, that is the variation of the flux of CRs on angular scales as small as $5^{\circ}$ which is absent in simple diffusion models. (See~\citet{Ahlers:2016rox} for a review on small-scale anisotropies).

What has been hampering progress are mainly two issues. First, the transport of high-energy, charged particles through a turbulent magnetised plasma is intrinsically non-linear: The temporal evolution of the phase space density of particles can be described by a Fokker-Planck equation with coefficients that depend on the small-scale magnetic field as will be reviewed below. At the same time, however, CRs contribute to the dielectric tensor of the plasma, thus affecting its dispersion relation. Broadly speaking, waves are damped if the phase space density is very isotropic, but they can grow if there is sufficient anisotropy\footnote{More precisely, a positive (negative) imaginary part of the wave frequency denotes a growing (damped) mode. The sign of the CRs' contribution to the imaginary part depends on the sign of the gradient of the cosmic ray density in momentum space along the so-called resonance circle, that is the contour along which CRs are resonantly scattering with waves of a certain frequency. See \citet{2005ppfa.book.....K} for a lucid account.}. In general, sources are distributed inhomogeneously, this leads to anisotropy in momentum and growth of wave modes. This is called the streaming instability and can lead to self-confinement of CRs. While this fact was known already in the 1960's~\citep{1969ApJ...156..445K,1971ApL.....8..189K,1975MNRAS.173..245S}, only recently has it been incorporated into (simple) phenomenological models~\citep{2012PhRvL.109f1101B,2018PhRvL.121b1102E}. Note that self-generated turbulence is also important close to the sources of Galactic CRs~\citep{Malkov:2012qd,Ptuskin:2008zz,Nava:2016szf,Nava:2019vvi}. The amplified magnetic fields necessary for shock acceleration to the highest energies are thought to be provided by a different, but related instability~\citep{2001MNRAS.321..433B}.

The other issue is the lack of a fundamental microscopic theory for the transport of charged particles through turbulent magnetic fields. More than 50 years since its inception, quasi-linear theory (QLT)~\citep{1966ApJ...146..480J,1966PhFl....9.2377K,1967PhFl...10.2620H,1970ApJ...162.1049H} is still very much the paradigm for phenomenological applications to Galactic cosmic rays. In QLT, the Fokker-Planck equation for the temporal evolution of the phase space density of CRs is derived in a perturbative approach where the force on a particle due to a turbulent magnetic field is evaluated along the unperturbed trajectory in a regular background field. The Fokker-Planck coefficients, most prominently the components of the spatial diffusion tensor, can be computed for a given model of turbulence, parametrised by the two-point function of the turbulent magnetic field. 
Famously, in QLT the interactions between plasma waves and particles are found to be resonant, meaning that particles of a certain gyroradius $r_{\text{g}} = v / \Omega$, $\Omega$ denoting the gyrofrequency, are only affected by waves with a wavenumber $k$ that satisfies $k r_{\text{g}} \mu \approx 1$ (for low-frequency waves like Alfv\'en waves) where $\mu$ is the cosine of the pitch-angle, that is the angle between the particle momentum $\vec{p}$ and the regular magnetic field $\Breg$, $\mu \equiv \vec{p} \cdot \Breg / (|\vec{p}| |\Breg|)$.

While some of QLT's predictions are \emph{qualitatively} confirmed by data, e.g.\ the rigidity-dependence\footnote{Rigidity $\mathcal{R}$ is defined as the ratio of particle momentum over electric charge, $\mathcal{R} = p c / (Z e)$. For ultra-relativistic particles, it is simply related to the energy $E$ and energy-per-nucleon $(E/A)$ as $\mathcal{R} = p c / (Z e) \simeq E / (Z e) = (A/Z) (E/A)/e$ with $A$ and $Z$ denoting the CR's mass and charge numbers, respectively.} of the diffusion coefficients, there are a number of concerns. The most famous one is the $90^{\circ}$ problem: Due to the resonance condition, particles with pitch-angle close to $90^{\circ}$ ($\mu \approx 0$) can only be in resonance with very large wavenumbers $k$ which for the usual turbulent spectra contain little energy. In the limit $\mu \to 0$, the scattering rate vanishes and particles cannot change direction along the background field resulting in ballistic transport. This is obviously at variance with the diffusive transport inferred from observations.

The root cause of the $90^{\circ}$ problem is the assumption of unperturbed trajectories in QLT. This is remedied in non-linear theories, where the decay of correlations leads to a broadening of the resonance condition which allows for efficient enough scattering through $90^{\circ}$. However, any such extension of QLT requires additional assumptions, for instance on the form of temporal decorrelation of the particle's trajectory. Ideally, one would test these non-linear theories by comparing their predictions with data from observations. While this approach is being followed by the heliospheric community, a difficulty remains in that the actual turbulence (if known) turns out to be much more complex than what is routinely assumed in analytical transport theories. Alternatively, transport theories can be tested by comparing their predictions with those from \emph{numerical} experiments.

Test particle simulations compute the transport of high-energy charged particles through prescribed electro-magnetic fields without taking into account the effect of the high-energy particles on the electro-magnetic fields. To this end, a realisation of the turbulent magnetic field is generated and the equations of motion (Newton-Lorentz equations) are solved for test particles, that is the contributions of the CR particles to the electromagnetic fields are ignored. Given the trajectories of a large enough number of test particles, one can numerically compute the Fokker-Planck coefficients.

This idea has been very popular ever since powerful enough computers have been available to allow for the computation of thousand if not millions of test particles. Yet, we have found the body of literature on this rather disjoint, with different groups employing incompatible prescriptions with no single widely agreed upon method for how to synthetically generate the turbulent magnetic field. It is the intention of this \emph{review} to provide a low-level introduction to the uninitiated while also discussing some of the applications of test particle simulations.

In addition to testing transport theories by comparing the analytically computed diffusion coefficients to simulated ones, there are at least two more applications of test particle simulations: First, for sources at distances closer or similar to the scattering mean free path, the diffusive transport theory is not necessarily applicable. An often-cited pathology of computing solutions to the diffusion equation is superluminal propagation speeds. Lately, there has been increased interest in the transition between the ballistic and diffusive phases of transport~\citep{2014ApJ...783...15E,2015ApJ...808..157M} and test particle simulations allow exploring this transition for given turbulent electro-magnetic fields (e.g.~\citealt{2016RAA....16..162T}). Second, analytical transport theories usually make predictions only for the ensemble-averaged phase-space density and it is usually assumed that the observed phase-space densities are close to the ensemble average. Recently, this has been called into question, in particular in view of the observation of small-scale anisotropies observed in the arrival directions of TeV-PeV CRs. Test particle simulations naturally simulate CR distributions for individual realisations of the turbulent fields and thus provide direct access to such stochasticity effects.

This \emph{review} will be structured as follows. In Sec.~\ref{sec:QLT} we give a brief review of QLT, describing how the diffusion coefficients are evaluated, introducing some of the simplest and most popular turbulence models. We will also review a few of QLT's non-linear extensions. In Sec.~\ref{sec:generating}, we explain the two main methods that have been employed in generating turbulent magnetic fields on a computer. We will reproduce the recipes from the literature in a way that should allow the interested reader to produce her/his own synthetic turbulence. In Sec.~\ref{sec:applications}, we will discuss two applications of test particle simulations, that is the computation of parallel and perpendicular mean free paths and the prediction of anisotropies in the arrival direction of CRs. Specifically, we will clarify some of the issues related to backtracking--a technique based on solving the equations of motion backward in time in Sec.~\ref{sec:validity_Liouville}. We will conclude with a short summary and outlook in Sec.~\ref{sec:conclusion}.

\section{Quasi-linear theory and extensions}
\label{sec:QLT}

For some 50 years, quasi-linear theory (QLT)~\citep{1966ApJ...146..480J,1966PhFl....9.2377K,1967PhFl...10.2620H,1970ApJ...162.1049H} has been the broadly accepted and widely employed theory of CR transport. Its success and popularity can be ascribed to its conceptual simplicity and validity in a number of important environments, including the solar wind, the interstellar medium and galaxy clusters. In addition, QLT is simple \emph{in principle} and thus allows for a straight-forward computation of the transport parameters, albeit it can become arbitrarily complex \emph{in practice}. Finally, these results can be found to agree with inferences from observations, e.g.\ the normalisation and power law shape of the Galactic diffusion coefficient.

At the heart of QLT is the evaluation of the turbulent magnetic field and its contribution to the Lorentz force along ``unperturbed orbits'', that is trajectories calculated in only a large-scale, regular magnetic field. Interactions of CRs with small-scale, magnetised turbulence result in resonant interactions, that is particles of Larmor radius $r_{\text{g}}$ and pitch-angle cosine $\mu$ interact predominantly with modes of wavenumber $k$ that satisfies $k r_{\text{g}} \mu \approx 1$. These resonant interactions lead to pitch-angle scattering and for a spectrum of magnetic turbulence with random phases, the particle performs a random walk in pitch-angle. The evolution of the phase space density can be described by a Fokker-Planck equation and the Fokker-Planck coefficients, e.g.\ the pitch-angle diffusion coefficient or the rate of second-order Fermi acceleration, depend on the two-point correlation functions of the turbulent magnetic field. In addition, under the assumption of slow variation of the phase space density with position and time, pitch-angle diffusion results in spatial diffusion along the background magnetic field~\citep{1988ApJ...331L..91E}. Finally, QLT also allows computing the dipole anisotropy in the arrival directions of CRs for a given spatial gradient of the phase space density.

In the following, we review the foundations of QLT, starting from the derivation of the Fokker-Planck equation. After an introduction to the various turbulence geometries in use, we outline how the transport coefficients can be computed. Motivated by the short-comings of QLT, we review some of its non-linear extensions.

\subsection{Derivation of the Fokker-Planck equation}
\label{sec:derivation_FPE}

Charged particles in electric and magnetic fields $\vec{E}$ and $\vec{B}$ are subject to the Lorentz force,
\begin{equation}
\vec{F}_{\text{L}} = e \left( \vec{E} + \vec{v} \times \vec{B} / c \right) \, ,
\label{eqn:Lorentz_force}
\end{equation}
with $e$ and $\vec{v}$ the charge and velocity of the particle and $c$ the speed of light. It is customary to decompose the magnetic field into a  large-scale, homogeneous, regular background field, $\Breg{}$ and a small-scale, turbulent, random field $\vec{\delta B}$, that is $\vec{B} = \Breg{} + \vec{\delta B}$ with $\avg{ \vec{\delta B} } \equiv 0$. (Throughout this article, we use angled brackets to denote averages over an ensemble of turbulent magnetic fields.) Without loss of generality, we assume in the following that the regular field is oriented along the $z$-direction, $\Breg{} = B_z \hat{z}$, unless stated otherwise. Large-scale electric fields are usually ignored, $\langle \vec{E} \rangle = 0$, as the large mobility of charges in astrophysical plasmas is efficiently shielding against regular electric fields (that is on scales much larger than the Debye length). Small-scale electric fields $\vec{\delta E}$ are necessarily present, but from Faraday's induction law, their magnitude can be estimated to be $|\vec{\delta E}| \sim (\vA / c) |\vec{\delta B}|$ with $\vA$ the Alfv\'en velocity and $\vA / c \ll 1$ in most astrophysical environments. Thus, to lowest order in $(\vA / c)$, there is no electric field and as the magnetic force is not performing any work on the particle, particle energy is consequently conserved. Note that at higher orders in $(\vA / c)$, the particle energy can change in a second-order Fermi type process. For simplicity, we constrain ourselves here to considering the lowest order case which results in pitch-angle scattering. For the fuller picture including the higher-order processes, we refer the interested reader to \citet{Schlickeiser:2002pg}.

A charged particle in a magnetic field forms a Hamiltonian system as long as dissipative processes (or any form of energy losses) can be ignored. A consequence of this is \emph{Liouville's theorem}, that is the conservation of phase space volume under canonical transformations. As time evolution is a canonical transformation, phases space volume is conserved in time~\citep{Goldstein-book}. Together with particle number conservation this implies the conservation of phase space \emph{density} $f = f(\vec{r}, \vec{p}, t)$. This is conveniently captured by what we will call \emph{Liouville's equation},
\begin{equation}
\frac{\dd f}{\dd t} = \frac{\partial f}{\partial t} + \frac{\dd \vec{r}}{\dd t} \cdot \vec{\nabla}_{\vec{r}} f + \frac{\dd \vec{p}}{\dd t} \cdot \vec{\nabla}_{\vec{p}} f = 0 \, ,
\label{eqn:Liouville_equation}
\end{equation}
encoding the incompressibility of the phase space flow. Here,
\begin{equation}
\frac{\dd \vec{r}}{\dd t} = \vec{v} \, , \quad \frac{\dd \vec{p}}{\dd t} = e \left( \vec{E} + \frac{\vec{v} \times \vec{B}}{c} \right) \, , \label{eqn:eqs_of_motion}
\end{equation}
are the equations of motion. Note that a necessary condition for a Hamiltonian system is that the forces are conservative and differentiable (``$p$-divergence-free''). 

A collisionless plasma under the influence of external electric and magnetic fields, $\vec{E}$ and $\vec{B}$, is an example of a Hamiltonian system. Its Hamiltonian is~\citep{Jackson:1998nia}
\begin{equation}
H = \sqrt{(c \vec{P} - e \vec{A})^2 + m^2 c^4} + e \Phi \, .
\end{equation}
Here, $\vec{P} = \vec{p} + (e/c) \vec{A}$ is the canonical momentum, $\vec{A}$ the vector potential, $m$ the particle mass, $e$ its charge and $\Phi$ the electric potential. Therefore, the phase space density of this collisionless plasma satisfies eq.~\eqref{eqn:Liouville_equation} and substituting the Lorentz force, eq.~\eqref{eqn:Lorentz_force}, in eq.~\eqref{eqn:Liouville_equation} gives the Vlasov equation,
\begin{equation}
\frac{\partial f}{\partial t} + \frac{\dd \vec{r}}{\dd t} \cdot \vec{\nabla}_{\vec{r}} f + e \left( \vec{E} + \frac{\vec{v} \times \vec{B}}{c} \right) \cdot \vec{\nabla}_{\vec{p}} f = 0 \, ,
\label{eqn:Vlasov}
\end{equation}
which together with Maxwell's equations forms the basis of plasma kinetic theory. For a \emph{collisional} plasma, a term needs to be added to the right-hand side, the famous collision operator. For a \emph{collisionless} plasma (as appropriate for CRs) the right-hand side remains zero.

Considering turbulent fields, the phase space density also becomes a random field, $f = \avg{ f } + \delta f$, with an expectation value, $\avg{ f }$, and fluctuations around it, $\delta f$, that satisfy $\avg{ \delta f } = 0$.

In any realistic astrophysical situation, it is of course impossible to know the small-scale turbulent field at all positions in order to exactly solve eq.~\eqref{eqn:Vlasov}. Instead, one can only hope to predict statistical moments of the phase space density for a statistical ensemble of turbulent magnetic fields. Traditionally, one is mostly interested in the first moment, the ensemble average, though see~\citet{Mertsch:2019xij} for the computation of a second-order moment.

In the following, we ignore electric fields, see above. Averaging eq.~\eqref{eqn:Vlasov}, we find, see e.g.~\citet{Jokipii1972},
\begin{align}
\frac{\dd \avg{f} }{\dd t} &= \frac{\partial \avg{f}}{\partial t} + \frac{\dd \vec{r}}{\dd t} \cdot \vec{\nabla}_{\vec{r}} \avg{f} + e \frac{\vec{v} \times \Breg{}}{c} \cdot \vec{\nabla}_{\vec{p}} \avg{f} \nonumber
\\ &= -\LRavg{ e \frac{\vec{v} \times \vec{\delta B}}{c} \cdot \nabla_{\vec{p}} \delta f } \neq 0 \, .
\label{eqn:averaged_Vlasov}
\end{align}
Note that unlike the phase space density $f$, the ensemble averaged phase space density $\avg{f}$ is \emph{not} conserved, $\dd \avg{f} / \dd t \neq 0$. (More on this in Sec.~\ref{sec:validity_Liouville}.)

One way to glean some physical insight from eq.~\eqref{eqn:averaged_Vlasov} is to identify its right-hand-side with a damping term,~\citep{1988ApJ...331L..91E,1989ApJ...340.1112W},
\begin{equation}
\LRavg{ e \frac{\vec{v} \times \vec{\delta B}}{c} \cdot \nabla_{\vec{p}} \delta f } \to \nu \left( \avg{f} - \frac{1}{4 \pi} \int \dd \vhat{p} \, \avg{f} \right) \, ,
\end{equation}
(where $\vhat{p} \equiv \vec{p} / |\vec{p}|$) that is driving the phase space density towards isotropy at a rate $\nu$, an approach that can also be motivated by gas kinetic theory~\citep{1954PhRv...94..511B}. This way, eq.~\eqref{eqn:averaged_Vlasov} can be solved and shown to lead to a spatial diffusion equation. The parallel diffusion coefficient can be identified as $\kappa_{\parallel} = v^2/(3 \nu)$ whereas the perpendicular diffusion coefficient satisfies $\kappa_{\perp}/\kappa_{\parallel} = (1 + \Omega^2/\nu^2)^{-1}$, a result referred to as the ``classical scattering limit''~\citep{1969P&SS...17...31G}. Here, $\Omega$ is the particle's gyrofrequency.

In QLT, however, a more systematic solution for $f$ is sought through an equation for the temporal evolution of the fluctuations $\delta f$. Such an equation can be obtained by subtracting the ensemble-averaged Vlasov eq.~(\ref{eqn:averaged_Vlasov}) from the original Vlasov eq.~(\ref{eqn:Vlasov}),
\begin{align}
& \frac{\partial \delta f}{\partial t} + \frac{\dd \vec{r}}{\dd t} \cdot \vec{\nabla}_{\vec{r}} \delta f + e \left( \frac{\vec{v} \times \Breg{}}{c} \right) \cdot \vec{\nabla}_{\vec{p}} \delta f \nonumber \\
& \simeq -e \left( \frac{\vec{v} \times \vec{\delta B}}{c} \right) \cdot \vec{\nabla}_{\vec{p}} \avg{f} \, . \label{eqn:deltaf_eqn}
\end{align}
Here, we have chosen to ignore the difference
\begin{equation}
e \frac{\vec{v} \times \vec{\delta B}}{c} \cdot \nabla_{\vec{p}} \delta f - \LRavg{ e \frac{\vec{v} \times \vec{\delta B}}{c} \cdot \nabla_{\vec{p}} \delta f } \, ,
\end{equation}
which is second order in perturbed quantities, $\vec{\delta B}$ and $\delta f$. This assumes, of course, that $|\vec{\delta B}| \ll |\Breg{}|$ and therefore $\delta f \ll \avg{f}$. Eq.~\eqref{eqn:deltaf_eqn} can now be integrated with the method of characteristics, the formal solution being
\begin{equation}
\delta f = \delta f_0 - \int_{t_0}^t \dd t' \left[ e \left( \frac{\vec{v} \times \vec{\delta B}}{c} \right) \cdot \vec{\nabla}_{\vec{p}} \avg{f} \right]_{P(t')} \, . \label{eqn:deltaf_sln}
\end{equation}
Here, $\delta f_0 \equiv \delta f(\vec{r}, \vec{p}, t_0)$ denotes the phase space density at time $t_0$ and the subscript $P(t')$ indicates that positions and momenta in the square brackets are to be evaluated along the characteristics of eq.~\eqref{eqn:deltaf_eqn}, that is the solutions of the equations of motions, eq.~\eqref{eqn:eqs_of_motion} with $\vec{B}$ replaced by the regular field $\Breg{}$ only (and again no electric field). These solutions $P$ are commonly referred to as ``unperturbed orbits'' or ``unperturbed trajectories''. For the homogeneous regular magnetic field $\Breg{} = B_z \hat{z}$ assumed here they are of course helices along the $z$-direction.

We can now substitute eq.~\eqref{eqn:deltaf_sln} into eq.~\eqref{eqn:averaged_Vlasov},
\begin{align}
& \frac{\partial \avg{f}}{\partial t} + \frac{\dd \vec{r}}{\dd t} \cdot \vec{\nabla}_{\vec{r}} \avg{f} + e \frac{\vec{v} \times \Breg{}}{c} \cdot \vec{\nabla}_{\vec{p}} \avg{f} \nonumber 
\\ &\simeq \int_{t_0}^t \!\! \dd t' \! \LRavg{ e \frac{\vec{v} \! \times \! \vec{\delta B}}{c} \! \cdot \! \nabla_{\vec{p}} \left[ e \frac{\vec{v} \! \times \! \vec{\delta B}}{c} \! \cdot \! \vec{\nabla}_{\vec{p}} \avg{f} \right]_{P(t')} } ,
\label{eqn:almost_Fokker_Planck}
\end{align}
where we have dropped the term $\propto \delta f_0$. At this stage, we can already see that the right-hand side will lead to diffusion terms (courtesy of the two momentum derivatives) and that it depends on the turbulent magnetic field's two-point function, integrated along the unperturbed trajectory $P(t')$. To make further progress, we consider the momentum $\vec{p}$ in spherical coordinates, that is $\vec{p} = p (\sqrt{1 - \mu^2} \cos\phi, \sqrt{1 - \mu^2} \sin\phi, \mu)^T$ and introduce the correlation lengths $l_{\text{c}}$ and correlation times $\tau_{\text{c}}$ of the turbulent magnetic field, defined through
\begin{align}
\avg{ \delta B^2 } l_{\text{c}} &\equiv \int_0^{\infty} \dd \Delta r \, \avg{\delta B(\vec{r}) \delta B(\vec{r}+\vec{\Delta r})} \, , \\
\avg{ \delta B^2 } \tau_{\text{c}} &\equiv \int_0^{\infty} \dd \Delta t \, \avg{\delta B(t) \delta B(t+\Delta t)} \, .
\end{align}
(Strictly speaking, the correlation lengths and times are tensors because of the vector nature of the magnetic field in the two-point functions; here, however, we only require them for order of magnitude arguments, so we do not distinguish the different components.) Here, $\delta B(t)$ is short-hand for $\left[ \delta B(\vec{r}) \right]_{P(t)}$, that is $\delta B(\vec{r})$ evaluated along the unperturbed trajectory at time $t$.

The right-hand side of eq.~\eqref{eqn:almost_Fokker_Planck} is still rather unwieldy and further progress requires a number of assumptions. In addition to
\begin{enumerate}
\item Smallness of perturbations, $|\vec{\delta B}| \ll |\Breg{}|$ (see above);
\end{enumerate}
these are:
\begin{enumerate}
\setcounter{enumi}{1}
\item Gyrotropy: The ensemble-averaged phase space density $\avg{f}$ does not depend on the azimuthal angle $\phi$, so $\avg{f} (\vec{r}, p, \mu, \phi, t) \to \avg{f} (\vec{r}, p, \mu, t)$.
\item Adiabatic approximation: The phase space density only varies on time-scales much larger than the correlation time of the turbulent magnetic field, $\tau_{\text{c}}$,
\begin{equation}
\avg{f} \left/ \frac{\partial \avg{f}}{\partial t} \right. \gg \tau_{\text{c}} \, .
\end{equation}
\item Finite correlation times: The correlation times of the turbulent magnetic field are much larger than the Larmor time, $\tau_{\text{c}} \gg \Omega^{-1}$.
\item Homogeneous and stationary turbulence. \label{item:homogeneous_stationary}
\end{enumerate}

Under these conditions, the ensemble averaged Vlasov equation ultimately results in a Fokker-Planck type equation~\citep{1914AnP...348..810F,1917AnP...358..241P}, also known as the Kolmogorov forward~\citep{Kolmogorov:1931} or as the Smoluchowski equation~\citep{Bogoliubov:1939}, describing diffusion in pitch-angle,
\begin{equation}
\frac{\partial \avg{f}}{\partial t} + v \mu \frac{\partial \avg{f}}{\partial z}  = \frac{\partial}{\partial \mu} \left( D_{\mu \mu} \frac{\partial \avg{f}}{\partial \mu} \right) \, .
\label{eqn:Fokker_Planck_simple}
\end{equation}
Following the approach sketched above, the pitch-angle diffusion coefficient,
\begin{equation}
D_{\mu\mu} \equiv \frac{\avg{(\Delta \mu)^2}}{2 \Delta t} \, , \label{eqn:definition_Dmumu}
\end{equation}
can be expressed in terms of the correlation function of the magnetic field.

If we had not decided to ignore any electric field, additional terms would have appeared in the Fokker-Planck equation~\eqref{eqn:Fokker_Planck_simple}, relating to changes in momentum $p$ and pitch-angle, with diffusion coefficients $D_{\mu p} = D_{p \mu}$ and $D_{pp}$ defined analogously to eq.~\eqref{eqn:definition_Dmumu}. We have furthermore assumed that $v_{\text{A}} / v \ll 1$ in order for $D_{xx}$, $D_{yy}$, $D_{xy}$ and $D_{yx}$ to be negligible. Not doing so, would have resulted in the additional terms
\begin{align}
& \frac{\partial}{\partial x} \left( D_{xx} \frac{\partial \avg{f}}{\partial x} + D_{xy} \frac{\partial \avg{f}}{\partial y} \right) \nonumber \\
+ & \frac{\partial}{\partial y} \left( D_{yx} \frac{\partial \avg{f}}{\partial x} + D_{yy} \frac{\partial \avg{f}}{\partial y} \right) \, ,
\end{align}
to be added to the right-hand side of eq.~\eqref{eqn:Fokker_Planck_simple}.

In summary, under the influence of a turbulent magnetic field, charged particles are performing a random walk in pitch-angle which in the ensemble average results in diffusion in pitch-angle (cosine).

\subsection{The diffusion approximation}
\label{sec:diffusion_approximation}

Particle transport can be conveniently categorised if the mean-square displacement in direction $i$, $\langle \Delta x_i^2 \rangle$, has a power law dependence,
\begin{equation}
\langle \Delta r_i^2 \rangle \propto (\Delta t)^{\alpha}
\end{equation}
as
\begin{equation}
\begin{array}{r l}
\alpha < 1: & \text{sub-diffusive,} \\
\alpha = 1: & \text{diffusive,} \\
\alpha > 1:& \text{super-diffusive, in particular} \\
\alpha = 2: & \text{ballistic.}
\end{array}
\end{equation}

It seems clear that transport in any perturbative theory with $|\vec{\delta B}| \ll |\Breg{}|$ must be ballistic at early enough times: Particles just gyrate around $\Breg{}$ and $\langle \Delta z^2 \rangle = (v \mu \Delta t)^2$ while $\langle \Delta x^2 \rangle = \langle \Delta y^2 \rangle = 0$ when integrated over full gyroperiods. At late times, that is for $t \gg D_{\mu\mu}^{-1}$, we would expect diffusive behaviour for the transport along the field.

In order to formalise this picture, we derive a spatial diffusion equation from the Fokker-Planck equation~\eqref{eqn:Fokker_Planck_simple}. To this end, we decompose $\avg{f}$ into an isotropic part, $g$, and an anisotropic part, $h$,
\begin{align}
\avg{f}(p, \mu, t) = g(p, t) + h(p, \mu, t) \, ,
\\ \text{where } g(p, t) = \frac{1}{2} \int_{-1}^1 \dd \mu \avg{f}(p, \mu, t) 
\\ \text{and} \quad \int_{-1}^1 \dd \mu \, h(p, \mu, t) = 0 \, .
\end{align}
If $g$ varies only slowly with time and position,
\begin{equation}
g \left/ \frac{\partial g}{\partial t} \right. \gg \tau_{\text{sc}} \quad \text{and} \quad g \left/ \frac{\partial g}{\partial z} \right. \gg \lambda_{\text{sc}} \, ,
\end{equation}
where $\tau_{\text{sc}} \sim D_{\mu\mu}^{-1}$ and $\lambda_{\text{sc}} \sim v \tau_{\text{sc}}$ are the scattering time and mean-free path, respectively, the phase space density will be very isotropic, $h \ll g$. In this case, we can derive a spatial diffusion equation for the isotropic part $g$~(e.g.~\citealt{1970ApJ...162.1049H}),
\begin{equation}
\frac{\partial g}{\partial t} - \frac{\partial}{\partial z} \left( \kappa_{\parallel} \frac{\partial g}{\partial z} \right) = 0 \, ,\label{eqn:diffusion_equation}
\end{equation}
with the parallel diffusion coefficient
\begin{equation}
\kappa_{\parallel} = \frac{v^2}{8} \int_{-1}^1 \dd \mu \frac{(1- \ \mu^2)^2}{D_{\mu\mu}} \, . \label{eqn:kappa_parallel}
\end{equation}
Furthermore, we would expect the anisotropic part $h$ to be dominated by the dipole anisotropy, that is $h \approx h_1 \mu$ with
\begin{equation}
h_1 = \frac{3}{2} \int_{-1}^1 \dd \mu \, \mu \, h(\mu) = -\frac{2}{v} \kappa_{\parallel} \frac{\partial g}{\partial z} \, .
\end{equation}

\subsection{Computation of transport coefficients}

So far, we have not specified the functional form of the Fokker-Planck coefficients, e.g.\ the pitch-angle diffusion coefficient $D_{\mu\mu}$, and its dependence on the two-point correlation function of turbulence $P_{ij}(\vec{k})$ that emerges in the derivation of the Fokker-Planck equation~\eqref{eqn:Fokker_Planck_simple}. An alternative to the derivation of Sec.~\ref{sec:derivation_FPE} is to directly compute the Fokker-Planck coefficients from solutions of the equations of motion. In fact, an arbitrary Fokker-Planck coefficient $D_{PQ}$ can be defined in terms of the mean displacements of the variables in question, $P$ and $Q$. For instance, the pitch-angle diffusion coefficient can be derived as the $t \to \infty$ limit of the running diffusion coefficient,
\begin{equation}
d_{\mu\mu}(t) = \frac{1}{2} \frac{\dd}{\dd t} \LRavg{ (\Delta \mu)^2 } \, . \label{eqn:running}
\end{equation}
This is a consequence of the Taylor-Green-Kubo formula~\citep{doi:10.1112/plms/s2-20.1.196,1951JChPh..19.1036G,1957JPSJ...12..570K}, 
\begin{equation}
D_{\mu\mu} = \int_0^{\infty} \dd t \avg{ \dot{\mu}(0) \dot{\mu}(t) } \label{eqn:Dmumu_TGK} \, ,
\end{equation}
where the dots denote derivatives with respect to time. For diffusive transport, eqs.~\eqref{eqn:definition_Dmumu} and \eqref{eqn:running} coincide, of course. Moreover, this allows computing the parallel diffusion coefficient $\kappa_{\parallel}$ without the detour of computing $D_{\mu\mu}$ first and then applying the diffusion approximation, eq.~\eqref{eqn:kappa_parallel}.

From the equations of motion, see eq.~\eqref{eqn:eqs_of_motion}, we find
\begin{equation}
\dot{\mu} = \frac{e}{c p} \left( \vec{v} \times \vec{B} \right)_z = \frac{1}{r_{\text{g}} \Bregular{}} \left(v_x \delta B_x(\vec{r}) - v_y \delta B_y(\vec{r}) \right) ,
\end{equation}
and thus
\begin{align}
& D_{\mu\mu} \nonumber \\ =& \frac{1}{\Bregular^2 r_{\text{g}}^2} \!\! \int_0^{\infty} \!\!\!\!\! \dd t \left[ v_x(t) v_x(0) \mathcal{P}_{yy}(t) \! + \! v_y(t) v_y(0) \mathcal{P}_{xx}(t) \right] . \label{eqn:Dmumu_1}
\end{align}
Here, we have defined
\begin{equation}
\mathcal{P}_{ij}(t) \equiv \avg{ \delta B_i(0) \delta B_j(t) } \, ,
\end{equation}
and both the velocities and the magnetic fields are to be evaluated along unperturbed trajectories. Note that the fact that the Fokker-Planck coefficients only depend on the two-point function means that we can constrain ourselves to the Gaussian part of the turbulent magnetic field.

\subsection{Turbulence geometries and spectra}
\label{sec:turbulence_models}

To make further progress, we need to specify the turbulence correlation tensor $P_{ij}$. In the derivation of the Fokker-Planck equation we had to assume that turbulence is homogeneous and stationary, that is its statistical moments are invariant under translations in space and time (see assumption~\ref{item:homogeneous_stationary}). In this case, the field can be represented very economically in Fourier space. To this end, we introduce the Fourier transform pair
\begin{align}
\delta \tilde{B}_j(\vec{k}, t) &= (2 \pi)^{-3/2} \int_{-\infty}^{\infty} \dd^3 r \, \delta B_j(\vec{r}, t) \ee^{\ii \vec{k} \cdot \vec{r}} \, , \label{eqn:FT_pair_1} \\
\delta B_j(\vec{x}, t) &= (2 \pi)^{-3/2} \int_{-\infty}^{\infty} \dd^3 k \, \delta \tilde{B}_j(\vec{k}, t) \ee^{-\ii \vec{k} \cdot \vec{r}} \, . \label{eqn:FT_pair_2} 
\end{align}
Note that for the magnetic field to have real values, $\delta B_j(\vec{r}) = \delta B^*_j(\vec{r})$, requires a relation between the Fourier components and their complex conjugates,
\begin{equation}
\delta \tilde{B}^*_j(\vec{k}) = \delta \tilde{B}_j(-\vec{k}) \, . \label{eqn:reality}
\end{equation}

The homogeneity and stationarity now guarantee that the two-point functions $\avg{ \delta B_i(\vec{r},t ) \delta B_j(\vec{r}', t) }$ depend on the positions $\vec{r}$ and $\vec{r}'$ and times $t$ and $t'$ only through the differences $\Delta \vec{r} \equiv (\vec{r} - \vec{r}')$ and $(t - t')$. It is then easy to see that the two-point function in Fourier space is diagonal,
\begin{align}
& \avg{ \delta\tilde{B}_i(\vec{k}, t) \delta\tilde{B}^*_j(\vec{k}', t') } \nonumber \\
&= \! (2 \pi)^{-3} \!\!\! \int \!\! \dd^3 r \, \ee^{\ii \vec{k} \cdot \vec{r}} \!\!\! \int \!\! \dd^3 r' \ee^{-\ii \vec{k}' \cdot \vec{r}'} \!\! \avg{ \delta B_i(\vec{r}, t) \delta B_j(\vec{r}'\!, t') } \\
&= \delta^{(3)} (\vec{k} - \vec{k}') P_{ij}(\vec{k}, t-t') \, ,
\end{align}
where the turbulence correlation tensor $P_{ij}(\vec{k}, \Delta t)$ is the Fourier transform of the coordinate space two-point function,
\begin{align}
P_{ij}(\vec{k}, \Delta t) \equiv & (2 \pi)^{-3/2} \int_{-\infty}^{\infty} \dd^3 (\Delta r) \, \ee^{\ii \vec{k} \cdot \Delta\vec{r}} \\
& \times \avg{ \delta B_i(\vec{r}, t) \delta B_j(\vec{r} - \Delta \vec{r}\!, t') } \, .
\end{align}
It contains all the (statistical) information on the magnetic turbulence that enters into the computation of the Fokker-Planck coefficients. This includes information on the turbulence geometry, for instance whether there is a preferred direction for the propagation of waves, information on the turbulence spectrum, that is the distribution of energy among different turbulent scales, as well as information on the time-dependence of the correlations. We will discuss a few parametrisations below.

Oftentimes, it is assumed that $P_{ij}(\vec{k}, \Delta t)$ factorises into a magnetostatic correlation tensor $P_{ij}(\vec{k}) \equiv P_{ij}(\vec{k}, 0)$ independent of time and a time-dependent dynamical correlation function $\Gamma(\vec{k}, \Delta t)$,
\begin{equation}
P_{ij}(\vec{k}, \Delta t) = P_{ij}(\vec{k}) \Gamma(\vec{k}, \Delta t) \, .
\end{equation}
In the magnetostatic approximation, we ignore any time-dependence altogether, so $\Gamma \equiv 1$.

While in reality $P_{ij}$ may be arbitrarily complicated, three turbulence geometries have dominated much of the literature, both in analytical studies of transport coefficients and numerical test particle simulations. These three geometries are conceptually simple and particularly amenable to analytical computations of  the components of the diffusion tensor and the other Fokker-Planck coefficients: 3D isotropic turbulence, slab turbulence and a composition of slab and 2D isotropic turbulence. In the following, we will give explicit formulas for the turbulence correlation tensor for these models in terms of a scalar power spectrum $g(k)$, the spectral part of the turbulence tensors. Afterwards, we introduce two popular parametrisations for $g(k)$ and conclude with an example for the computation of the pitch-angle diffusion coefficient.

\subsubsection{3D isotropic turbulence}
\label{sec:3D_isotropic_turbulence}

It is easy to show~\citep{1982tht..book.....B} that for 3D isotropic turbulence the magnetostatic correlation tensor takes the form
\begin{equation}
P^{\text{3D}}_{ij}(\vec{k}) = g^{\text{3D}}(k) \left( \delta_{ij} - \frac{k_i k_j}{k^2} + \ii \sigma(k) \epsilon_{ijm} \frac{k_m}{k} \right) \, ,
\end{equation}
with $k = |\vec{k}|$. The $k$-dependent real functions $g^{\text{3D}}(k)$ and $\sigma(k)$ allow modelling of the overall spectrum and of a wavenumber-dependent helicity, respectively. Note that for linearly polarised waves $\sigma(k) \equiv 0$. The normalisation of $g^{\text{3D}}(k)$ is fixed by requiring
\begin{align}
\delta B^2 &\equiv \avg{ \vec{\delta B}^2(\vec{x}) } \! = \!\! \int \dd^3 k \left( P^{\text{3D}}_{xx}(\vec{k}) + P^{\text{3D}}_{yy}(\vec{k}) + P^{\text{3D}}_{zz}(\vec{k}) \right) \nonumber \\
&= 2 \int \dd^3 k \, g^{\text{3D}}(k) = 8 \pi \int_0^{\infty} \dd k \, k^2 g^{\text{3D}}(k) \, .
\end{align}

\subsubsection{Slab turbulence}

In slab turbulence, it is assumed that all quantities are independent of the coordinates perpendicular to the background field (in our case: $x$ and $y$) and that the turbulent field has no $z$-component. Consequently, the wave vectors $\vec{k} \parallel \vhat{z}$ and if we further demand turbulence to be axisymmetric, the turbulence correlation tensor reads
\begin{equation}
P^{\text{slab}}_{ij}(\vec{k}) = g^{\text{slab}}(k_{\parallel}) \frac{\delta(k_{\perp})}{k_{\perp}} \left( \delta_{ij} + \ii \sigma(k_{\parallel}) \epsilon_{ijz} \right) \, ,
\label{eqn:correlation_tensor_slab}
\end{equation}
for $i, j \in x, y$ and zero otherwise. In our case, $k_{\parallel} = k_z$ and $k_{\perp} = \sqrt{k_x^2 + k_y^2}$. Again, $\sigma(k_{\parallel})$ allows for wavenumber-dependent helicity, but vanishes for linear polarisation. The normalisation is then
\begin{align}
\delta B^2 &\equiv \int \dd^3 k \left( P^{\text{slab}}_{xx}(\vec{k}) + P^{\text{slab}}_{yy}(\vec{k}) \right) \\*
&= 4 \pi \int_{-\infty}^{\infty} \dd k_{\parallel} \, g^{\text{slab}}(k_{\parallel}) \, . \label{eqn:slab_normalisation}
\end{align}

While slab turbulence might seem rather restrictive a turbulence model, it is quite attractive due to its simplicity. In addition, it could be argued that it is of physical relevance in situations where the turbulence is self-generated by anisotropies in the distribution of CRs \citep{1969ApJ...156..445K,1975MNRAS.173..245S}: It has been shown (e.g.~\citealt{1969ApJ...158..959T}) that the modes with wavevectors along the background magnetic field grow fastest.

\subsubsection{Composite (slab + 2D isotropic) turbulence}

Motivated by observations of the turbulence in the solar wind \citep{1990JGR....9520673M}, the heliospheric community has adopted a composite model for the correlation tensor as a superposition of a slab component and a 2D isotropic component. The motivation for this composite turbulence model were observations of CR mean-free paths which were in conflict with the observed turbulent energy densities. In fact, the observed mean-free path was significantly larger than what was predicted for the measured turbulence level in a pure slab model. As 2D turbulence contributes to pitch-angle scattering (and therefore to the parallel mean-free path) only marginally, moving part of the turbulent energy density from the slab to the 2D component, the measured level of turbulence could be reconciled with the mean-free path. According to \citet{1994ApJ...420..294B}, a $80 \, \%$ to $20 \, \%$ split between 2D and slab turbulence, respectively, reconciles the available data sets.

For linearly polarised waves, we can write
\begin{equation}
P^{\text{comp}}_{ij}(\vec{k}) = P^{\text{slab}}_{ij}(\vec{k}) + P^{\text{2D}}_{ij}(\vec{k}) \, ,
\end{equation}
with $P^{\text{slab}}_{ij}(\vec{k})$ as in eq.~(\ref{eqn:correlation_tensor_slab}) and
\begin{equation}
P^{\text{2D}}_{ij}(\vec{k}) = g^{\text{2D}}(k_{\perp}) \frac{\delta(k_{\parallel})}{k_{\perp}} \left( \delta_{ij} - \frac{k_i k_j}{k^2} \right) \, ,
\end{equation}
for $i, j \in x, y$ and zero otherwise. This turbulent 2D field only depends on the $x$- and $y$-coordinate, and has no $z$-component. The normalisation condition for the 2D component is
\begin{align}
\delta B^2_{\text{2D}} &\equiv \avg{ \vec{\delta B}^2(\vec{x}) } \! = \!\! \int \dd^3 k \left( P^{\text{2D}}_{xx}(\vec{k}) + P^{\text{2D}}_{yy}(\vec{k}) \right) \nonumber \\
&= 2 \pi \int \dd k_{\perp} \, g^{\text{2D}}(k_{\perp}) \, .
\end{align}

\subsubsection{Turbulence spectra}

Having reviewed three simple turbulence geometries, we need to specify the spectral shapes $g(k)$ in order to compute transport coefficients. In cascade models of turbulence~\citep{1941DoSSR..30..301K,1963AZh....40..742I,1965PhFl....8.1385K}, energy is injected on the largest scales in the so-called energy range. Non-linear interactions transfer energy to smaller scales over the so-called inertial range. At very small scales, the turbulent energy is dissipated in the so-called dissipation range. The scale separating the energy and inertial ranges is called the outer scale of turbulence and the scale  separating the inertial and the dissipation range is called the dissipation scale. For an introduction to turbulence theory, see e.g.~\citet{1995tlan.book.....F}. Both turbulence theory and observations point at the existence of power law spectra in the inertial range. In fact, power law spectra have been observed in interplanetary and interstellar space~\citep{1995ApJ...443..209A}. (For a review on interstellar turbulence, see~\citealt{2004ARA&A..42..211E}).

Both in numerical simulations and in analytical work, most authors have confined themselves to one of two spectra. The first one is a simple power law with spectral index $q$ and low wavenumber cut-off $k_0$, corresponding to the outer scale $(2 \pi / k_0)$,
\begin{equation}
g_{\text{PL}}(k) = \left\{ \begin{array}{ll} g_0 (k/k_0)^{-q} & \text{for } k \geq k_0 \, , \\ 0 & \text{otherwise.} \end{array} \right. \label{eqn:PL}
\end{equation}
The alternative is a broken power law with a flat spectrum below the wavenumber $k_0$ and a power law slope $q$ above,
\begin{equation}
g_{\text{BPL}}(k) = g_0 \left( 1 + \left( \frac{k}{k_0} \right)^{1/s} \right)^{-q s} \, . \label{eqn:BPL}
\end{equation}
Here, $s$ is parametrising the softness of the break and $s \to 0$ corresponds to a sharp break. It is assumed that the broken power law form can potentially also capture turbulence in the energy range, that is for $k < k_0$.

\subsubsection{Slab turbulence with broken power law spectrum}

By ways of example, we report the result for the pitch-angle diffusion coefficient in slab turbulence and for the broken power law spectrum~\citep{Shalchi_book},
\begin{equation}
g^{\text{slab}}(k_{\parallel}) = \frac{C \left(q, \frac{1}{2} \right)}{2 \pi k_0} \delta B^2 \left( 1 + \left( \frac{k}{k_0} \right)^2 \right)^{-q/2} \, .\label{eqn:gslab}
\end{equation}
The function $C(q, s)$ is fixed by the normalisation condition, see eq.~\eqref{eqn:slab_normalisation},
\begin{align}
\frac{1}{C(q, s)} & \equiv \frac{4}{k_0} \int_0^{\infty} \dd k \, \left( 1 + \left( \frac{k}{k_0} \right)^{1/s} \right)^{-q s} \nonumber 
\\ &= \frac{4}{k_0} \frac{\Gamma(s(q-1)) \Gamma(1+s)}{\Gamma(q \, s)} \, , \label{eqn:normalisation}
\end{align}
where $\Gamma(\cdot)$ denotes the gamma function. We have assumed that $q>1$ in order for the $k$-integral in eq.~\eqref{eqn:normalisation} to converge. For $q=1$, instead, we need to assume a cut-off, that is $g^{\text{slab}} = 0$ for $k_{\parallel} > k_{\text{max}}$.

Substituting eq.~\eqref{eqn:gslab} into eqs.~\eqref{eqn:correlation_tensor_slab} and \eqref{eqn:Dmumu_1}, one encouters the resonance function
\begin{equation}
R^{\text{slab}} = \pi \delta(k_{\parallel} \mu v \pm \Omega) \, , \label{eqn:slab_resonance_function}
\end{equation}
see \citet{Schlickeiser:2002pg} for details. Eventually, this simplifies to
\begin{equation}
D_{\mu\mu} = \frac{\pi}{2} C \! \left( \! q, \frac{1}{2} \! \right) q k_0 \frac{\delta B^2}{B_z^2} \frac{(1 - \mu^2) \mu^{q-1} (r_{\text{g}} k_0)^{q-2}}{(1 + \mu^2 (r_{\text{g}} k_0))^{q/2}} . \label{eqn:Dmumu_in_slab_turbulence}
\end{equation}
Here, $r_{\text{g}}$ denotes again the particle's gyroradius.

For relativistic particles $r_{\text{g}} \propto \mathcal{R}$ (where $\mathcal{R}$ again denotes rigidity) and if $\mathcal{R}$ is small enough such that $\mu^2 (r_{\text{g}} k_0) \ll 1$, we observe that the rigidity-dependence of $D_{\mu\mu}$ is of power law form reflecting the power law nature of the underlying turbulence spectrum. For Kolmogorov and Kraichnan type values, $q = 5/3$ and $3/2$, the rigidity-dependence of the pitch-angle diffusion coefficient is $D_{\mu\mu} \propto \mathcal{R}^{-1/3}$ and $\mathcal{R}^{-1/2}$ and the spatial diffusion coefficient $\kappa_{\parallel} \sim 1 / D_{\mu\mu} \propto \mathcal{R}^{1/3}$ and $\mathcal{R}^{1/2}$, respectively.

\subsection{Field-line random walk}
\label{sec:FLRW}

The computation of the pitch-angle diffusion coefficient in eqs.~\eqref{eqn:Dmumu_TGK}, \eqref{eqn:Dmumu_1} and \eqref{eqn:Dmumu_in_slab_turbulence} is based on an evaluation of the turbulent part of the Lorentz force along trajectories around the homogeneous background field. As long as perturbations are small, this gives the dominant contribution to the parallel diffusion coefficient, eq.~\eqref{eqn:kappa_parallel}. 

For perpendicular transport, however, there is another important contribution due to the fact that the field line is not perfectly homogeneous. Instead, the large-scale magnetic field evaluated for a particle along a field line changes direction with distance along this field line. Under certain conditions, this movement can be shown to be diffusive, see below. If the movement of the particle due to this effect is included in the computation of the mean-square displacements (or equivalently through the Taylor-Green-Kubo approach), this gives the so-called field-line random walk (FLRW) contribution to perpendicular transport. The contribution without this is oftentimes called the microscopic contribution.

For slab turbulence, the microscopic diffusion coefficient vanishes (the transport is in fact sub-diffusive), hence FLRW gives the only contribution. For other turbulence geometries, FLRW can also contribute, but might not be dominating.

Let's again assume the regular background field $\avg{ \vec{B} } = B_z \hat{z}$ to be dominating over the perturbations $\vec{\delta B}$. The equation determining the field line $\{ x(z), y(z) \}$ is
\begin{equation}
\frac{\dd x}{\dd z} = \frac{\delta B_x}{B_z} \, ,
\label{eqn:field_line}
\end{equation}
and similarly for $y(z)$. This can formally be integrated to obtain the mean square displacement in the perpendicular directions, e.g.
\begin{equation}
\avg{ (\Delta x)^2 } \! = \! \frac{1}{B_z^2} \int_0^z \!\!\! \dd z' \!\! \int_0^z \!\!\! \dd z'' \avg{ \delta B_x(\vec{r}(z')) \delta B_x(\vec{r}(z'')) } . \label{eqn:integral_FLRW}
\end{equation}
In slab turbulence, the integrand only depends on $z$ and it is easy to show that the perpendicular mean-square displacement $\avg{ (\Delta r_\perp)^2 }$ is ballistic at small $z$ and diffusive for large $z$, e.g.
\begin{equation}
\avg{ (\Delta x)^2 } = \left\{ \begin{array}{l l}
z^2 \left( \delta B_x / B_z \right)^2 & \text{for } z \to 0 \, , \\
2 \kappa_{\text{FLRW}} |z| & \text{for } z \to \infty \, ,
\end{array} \right.
\end{equation}
with the FLRW diffusion coefficient
\begin{equation}
\kappa_{\text{FLRW}} = \frac{2 \pi^2}{B_z^2} g^{\text{slab}}(0) \, .
\end{equation}

In other turbulence geometries, the integrand in eq.~\eqref{eqn:integral_FLRW} also depends on $x$ and $y$, such that an explicit solution is not possible without further assumptions. See \citet{Shalchi_book} for a more detailed discussion.

If particles are assumed to diffuse along field lines, $\avg{ (\Delta z)^2 } \propto \Delta t$, FLRW leads to subdiffusive perpendicular transport, $\avg{ (\Delta r_{\perp})^2 } \propto \sqrt{\Delta t}$, a phenomenon known as compound (sub)diffusion~\citep{1966ApJ...146..480J,1995PhRvL..75.2136M,2006ApJ...644..622R,2006ApJ...644..971R}. Theoretical predictions~\citep{2000ApJ...531.1067K} have been largely confirmed by numerical simulations~\citep{Giacalone:1999,2000ApJ...538..192M,2002GeoRL..29.1048Q}. Compound subdiffusion has been applied to a variety of environments like laboratory plasmas~\citep{1978PhRvL..40...38R,1991PPCF...33..795I}, the heliosphere~\citep{1969ApJ...155..777J,2006ApJ...639L..91Z}, Galactic transport~\citep{1963SvA.....6..477G,1971ApL.....8...93L,1993A&A...279..278C}, near-source transport~\citep{2013MNRAS.429.1643N}
 as well as shock acceleration~\citep{1994A&A...285..687A,1995A&A...302L..21D,1996A&A...314.1010K}.

\subsection{Short-comings of QLT}

Despite its popularity, QLT exhibits a number of issues which we will briefly review in the following.

The most well-known pathology of magnetostatic QLT is its inability to scatter particles through $90^{\circ}$. While present in a number of turbulence geometries, it is easiest illustrated in slab turbulence where the dependence of the pitch-angle diffusion coefficient $D_{\mu\mu}$ on the spectrum $g^{\text{slab}}(k)$ becomes very simple. In fact, inspecting eq.~\eqref{eqn:Dmumu_in_slab_turbulence} we see that $D_{\mu\mu} \to 0$ for $\mu \to 0$.

The root cause for the $90^{\circ}$ problem is the narrow resonance condition in magnetostatic QLT, $k_{\parallel} \mu \, r_{\text{g}} = \pm 1$, see eq.~\eqref{eqn:slab_resonance_function}. Particles at finite $\mu$ are in resonance with waves of finite parallel wavenumber, $k_{\parallel} = \pm 1 / (\mu r_{\text{g}})$. For $\mu$ approaching $0$, however, the resonant parallel wavenumber grows without bounds. With the turbulence spectra being falling power laws, however, there is only little energy at small scales and the pitch-angle scattering rate vanishes. In practice, there is of course no energy at all at scales below the dissipation scale.

We note that the vanishing of $D_{\mu\mu}$ does not necessarily imply that the parallel diffusion coefficient $\kappa_{\parallel}$ diverges. In fact, for slab turbulence, the $\mu$-integral in eq.~\eqref{eqn:kappa_parallel} remains finite as long as $q<2$. Whether the QLT prediction of $D_{\mu\mu}$ near $\mu=0$ and of $\kappa_{\parallel}$ are accurate is a different question altogether; test particle simulations can provide answers. We also note that for $q=1$ it might appear from eq.~\eqref{eqn:Dmumu_in_slab_turbulence} that there is no $90^{\circ}$ problem, however, eq.~\eqref{eqn:Dmumu_in_slab_turbulence} was derived under the assumption of $q>1$. In fact, for $q=1$, the necessary cut-off in $g^{\text{slab}}$ leads to a finite resonance gap for $|\mu| < \Omega / (v k_{\parallel})$. Finally, for isotropic turbulence, $D_{\mu\mu}$ also vanishes at $\mu=0$ and this time also $\kappa_{\parallel}$ diverges, even for $q<2$~\citep{2006JPhG...32..809T}. This is in stark contrast with test particle simulations which have shown that parallel transport is perfectly diffusive in isotropic turbulence, meaning that $\kappa_{\parallel}$ attains a finite value.

Another comment is in order: In the above discussion, we have constrained ourselves to the simplest turbulence model, in particular slab turbulence, linear polarisation, and considered the limit $\vA/v \to 0$ where $v$ is again the particle speed. If we had allowed for oblique propagation of waves, we would have had to deal with compressive modes, like the magnetosonic mode. Due to its finite $\delta B_z$ component, the magnetosonic wave allows for transit-time damping, that is another resonant interaction besides gyro-resonance. Note however, that this does not cure the $90^{\circ}$ problem. It can be shown that the gyro-resonant interactions to $D_{\mu\mu}(\mu=0)$ is $\propto (\vA/v)^q$ while the contribution from transit-time damping is vanishing at $\mu=0$. The parallel diffusion coefficient in turn is determined solely by the gyro-resonant contribution and scales like $(\vA/v)^{1-q}$, thus again diverging in the limit of $\vA/v \to 0$.

Nature has of course no difficulty to scatter particles through $90^{\circ}$, as evidenced by the isotropy of Galactic CRs. Therefore, the vanishing of $D_{\mu\mu}$ at $\mu = 0$ must be considered a theoretical issue. It was realised early on~(\citealt{1975RvGSP..13..547V}, see cf.~\cite{2008ApJ...685L.165T} for other references) that the origin of the $90^{\circ}$ problem is actually the delta-like resonance function of QLT in the magnetostatic approximation and it was claimed that plasma wave effects or dynamical turbulence would in fact cure this issue. Other authors~\citep{2006JPhG...32.1045T} have however pointed out that non-linear effects are likely more important. Non-linear theories, in particular, exhibit finite resonance widths, thus curing the $90^{\circ}$ problem. In addition, non-resonant scattering can also play a role, e.g.~\citet{1999ApJ...518..974R}. A certain degree of non-resonant scattering has been inferred from PIC simulations of Whistler wave turbulence~\citep{2015GeoRL..42.3114C}, for instance.

\begin{figure}[!tb]
\includegraphics[width=\columnwidth,trim={0.cm 3.cm 0.cm 0.cm}, clip=false]{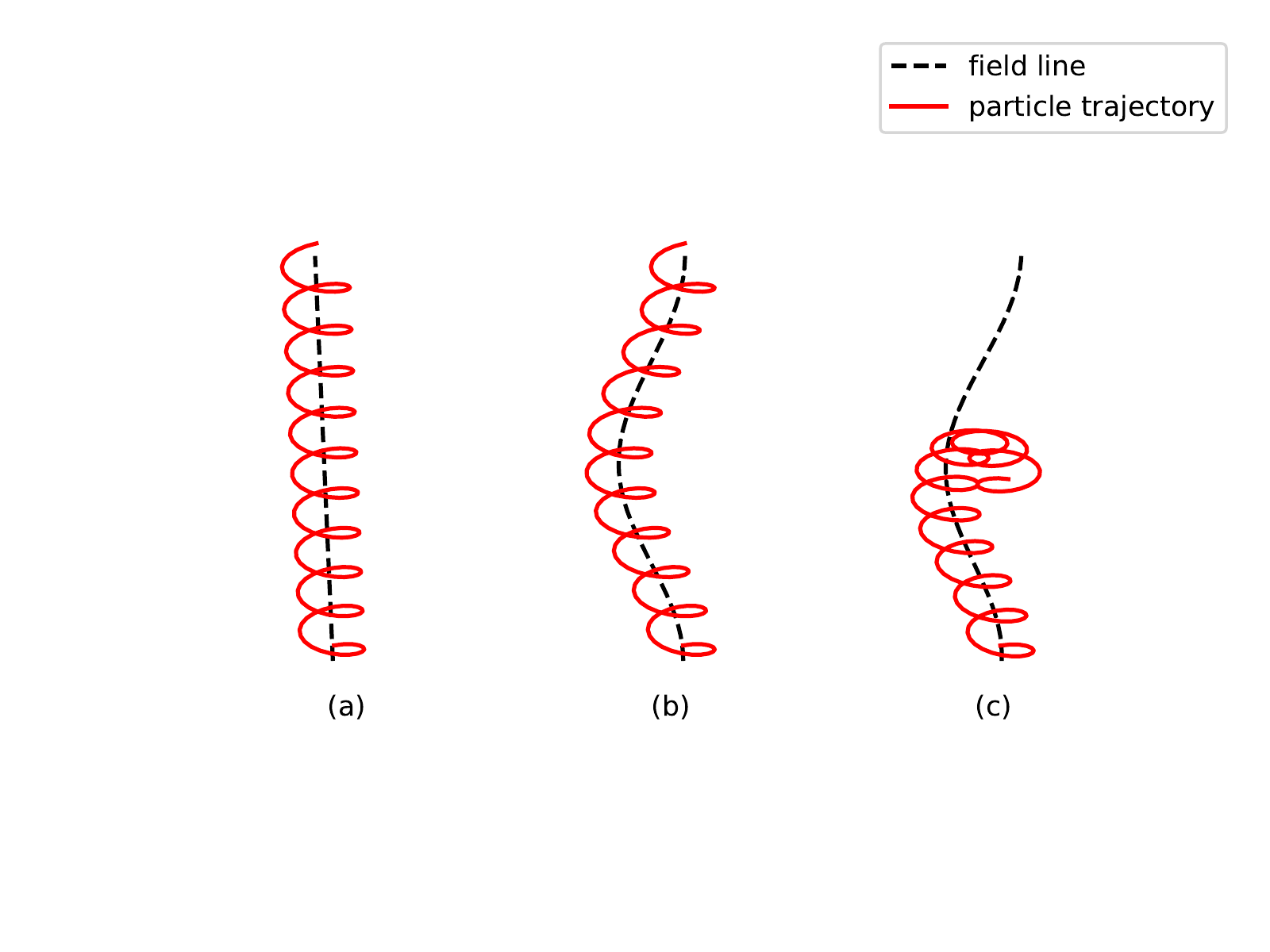}
\caption{Illustration of the different choices for how to evaluate perpendicular velocities when considering the perpendicular transport: (a) standard quasi-linear theory; (b) quasi-linear theory with field line random walk (FLRW); (c) FLRW and diffusion along the field line. See text for details.}
\label{fig:perpendicular_transport}
\end{figure}

Another important issue with QLT is its difficulty in describing perpendicular transport for slab turbulence. Whereas simulations find subdiffusive behaviour, $\avg{ (\Delta r_{\perp})^2 } \propto \sqrt{\Delta t}$~\citep{2002GeoRL..29.1048Q}, the answer from analytical models is not quite as clear and depends on what kind of assumptions enter the definition of the perpendicular displacements and which equations of motion are assumed. If we define the perpendicular diffusion coefficients as found in the derivation of the Fokker-Planck equation~\eqref{eqn:Fokker_Planck_simple}, we assume the equations of motion as in eq.~\eqref{eqn:eqs_of_motion}, meaning that the turbulent field is evaluated along the unperturbed trajectories in the homogeneous background field $\avg{ \vec{B}}$, see Fig.~\ref{fig:perpendicular_transport}a. In this case, $\kappa_{\perp}$ vanishes~\citep{Schlickeiser:2002pg}, again due to the narrow resonance condition. This assumption is of course strictly only true for small enough turbulent magnetic fields. If we instead make the assumption that particles follow field lines, see Fig.~\ref{fig:perpendicular_transport}b, diffusive behaviour is found, $\avg{ (\Delta r_{\perp})^2 } \propto \Delta t$. However, what has been ignored here is the diffusive nature of transport \emph{along} the field line. If this is taken into account, see Fig.~\ref{fig:perpendicular_transport}c, subdiffusive behaviour is found again. (This is the compound subdiffusion of Sec.~\ref{sec:FLRW} above.) Numerical simulations indeed confirm the subdiffusive behaviour. Whether the ambiguity of evaluating the perpendicular transport is an issue with QLT or of the additional assumptions made when evaluating $\avg{ (\Delta r_{\perp})^2 }$ is a matter of debate. Note that for non-slab geometries, diffusive behaviour is recovered.

Finally, it has been noted~\citep{Shalchi_book} that in other turbulence geometries there are also deviations between the QLT predictions and numerical results. Noteworthy are the deviations for composite geometry~\citep{2004ApJ...616..617S}.

\subsection{Non-linear extensions}
\label{sec:extensions}

So far, we have only considered magnetostatic turbulence which for QLT implies the $\delta$-like resonance function. Both dynamical turbulence and plasma wave damping lead to broadening of the resonance function. This has the potential of curing some of the deficiencies of QLT. (See~\citet{2006JPhG...32.1045T} for a discussion of the failure of QLT in \emph{undamped} plasma wave models.)

Another way to broaden the resonance function are non-linear theories. These replace the unperturbed orbits of QLT with perturbed orbits, that are more realistic at finite turbulence levels. Below, we review a number of non-linear theories and cite their respective resonance functions.

\subsubsection{BAM model~\citep{BAM-model}}

\citet{BAM-model} start from the velocity autocorrelation functions, $V_{ij}(t) \equiv \avg{ v_i(0) v_j(t) }$ that are required for computing diffusion coefficients with the TGK formalism,
\begin{equation}
\kappa_{ij} = \int_0^{\infty} \dd t \, \avg{ v_i(0) v_j(t) } \, .
\label{eqn:kappa_ij_TGK}
\end{equation}

In QLT, particle trajectories are perfect helices and the velocities of a particle along its trajectory stay correlated forever. This leads to simple, oscillatory correlations, $V_{xx}(t) = V_{yy}(t) \propto \cos \Omega t$ and $-V_{xy}(t) = V_{yx}(t) \propto \sin \Omega t$. In reality, however, velocities will not stay correlated indefinitely as particles will scatter in pitch-angle, and therefore these correlations should decay with time. In the BAM model, the decay is assumed exponential and thus the velocity correlation functions read
\begin{align}
V_{xx}(t) = V_{yy}(t) &= \frac{v^2}{3} \cos \Omega t \exp [- \omega_{\perp} t ] \, , \label{eqn:BAM_Vxx}
\\ -V_{xy}(t) = V_{yx}(t) &= \frac{v^2}{3} \sin \Omega t \exp [- \omega_{\text{A}} t ] \, , \label{eqn:BAM_Vyy}
\\ V_{zz}(t) &= \frac{v^2}{3} \exp [- \omega_{\parallel} t ] \, . \label{eqn:BAM_Vzz}
\end{align}

Substituting those into eq.~(\ref{eqn:kappa_ij_TGK}), one finds for the diffusion coefficients,
\begin{align}
\kappa_{\perp} = \kappa_{xx} = \kappa_{yy} &= \frac{v^2}{3} \frac{\omega_{\perp}}{\omega_{\perp}^2 + \Omega^2} \, ,
\\ \kappa_{\text{A}} = -\kappa_{xy} = \kappa_{yx} &= \frac{v^2}{3} \frac{\Omega}{\omega_{\perp}^2 + \Omega^2} \, ,
\\ \kappa_{\parallel} = \kappa_{zz} &= \frac{v^2}{3} \frac{1}{ \omega_{\parallel} } \, .
\end{align}
This is of similar form as the classical classical scattering result~\citep{1969P&SS...17...31G}, see Sec.~\ref{sec:derivation_FPE}.

In order to fix the perpendicular decorrelation rate, \citet{BAM-model} consider FLRW and postulate that the distance $z_c$ over which the field lines decorrelate is $z_c = r_{\text{g}}^2 / \kappa_{\text{FLRW}}$ and thus $\omega_{\perp} = v / z_c = v \kappa_{\text{FLRW}} / r_{\text{g}}^2$.

For a given turbulence geometry and spectrum, both $\kappa_{\parallel}$ and $\kappa_{\text{FLRW}}$ can be computed and the BAM model then allows determining $\kappa_{\perp}$ and $\kappa_{\text{A}}$. In slab turbulence, however, the BAM model predicts diffusive behaviour in the perpendicular direction which is at variance with what is seen in simulations. Furthermore, in composite turbulence (slab+2D) the BAM model cannot deal with the superdiffusive behavior of FLRW seen in simulations~\citep{Shalchi_book}. We thus conclude that the BAM model does not agree with simulation results, at least for two of the most important turbulence geometries.

\subsubsection{Non-linear guiding centre (NLGC) theory}

Non-linear guiding centre (NLGC) theory~\citep{2003ApJ...590L..53M} improves upon the velocity correlation functions of the BAM model insofar as that the perpendicular velocities ($i \in x, y$) are assumed to fulfill
\begin{equation}
v_i = a v_z \frac{\delta B_i}{\delta B_z} \, , \label{eqn:NLGC_idea}
\end{equation}
where $a$ is a free parameter that needs to be determined by fitting to simulations. This is inspired by the requirement for particle guiding centres to stay on field lines. In fact, for $a=1$, eq.~\eqref{eqn:NLGC_idea} reduces to the field line equation~\eqref{eqn:field_line}.

The perpendicular diffusion coefficient is then evaluated with the Taylor-Green-Kubo formula~\citep{doi:10.1112/plms/s2-20.1.196,1951JChPh..19.1036G,1957JPSJ...12..570K}. This gives four-point correlation functions $\avg{ v_z(0) v_z(t) \delta B_x(0) \delta B_x(t) }$ with two factors of magnetic field strength and two factors of parallel velocity. In NLGC theory this is assumed to factorise into two two-point functions. The (parallel) velocity part has a simple exponential form, if pitch-angle diffusion is isotropic, i.e.\ $D_{\mu\mu} \propto (1 - \mu^2)$. The Fourier transform of the two-point correlation for the magnetic field is further assumed to factorise into the power spectrum $P_{xx}$ and the so-called characteristic function, $\avg{ \exp [\ii \vec{k} \cdot \vec{\Delta x} ] }$. If the particle separations $\vec{\Delta x}$ are assumed normal-distributed and diffusive, e.g.\ $\avg{ (\Delta x)^2 } = 2 \kappa_{\perp} t$, the characteristic function takes a simple Gaussian form and the perpendicular diffusion coefficient reads,
\begin{align}
\kappa_{\perp} &= \frac{a^2}{B_z^2} \frac{v^2}{3} \int \dd^3 k \int_0^{\infty} \dd t \, P_{\perp}(\vec{k}, t) \nonumber
\\ & \times \exp [ - \omega_{\parallel} t - \kappa_{\perp} k_{\perp}^2 t - \kappa_{\parallel} k_{\parallel}^2 t ] \, .
\end{align}
With a power spectrum of the form $P_{\perp}(\vec{k}, t) = P_{\perp}(\vec{k}) \Gamma(\vec{k}, t)$ and a dynamical correlation function $\Gamma(\vec{k}, t) = \exp [ - \gamma(\vec{k}) t ]$ this simplifies to
\begin{equation}
\kappa_{\perp} = \frac{a^2}{B_z^2} \frac{v^2}{3} \int \dd^3 k \frac{P_{\perp}(\vec{k})}{ \omega_{\parallel} + \kappa_{\perp} k_{\perp}^2 + \kappa_{\parallel} k_{\parallel}^2 + \gamma(\vec{k}) } \, .
\label{eqn:NLGC_integral}
\end{equation}
Note how the sought-for perpendicular diffusion coefficient appears on both sides of the equations. Oftentimes, $\kperp$ is therefore computed iteratively.

For slab turbulence and in the magnetostatic case ($\gamma=0$), the integral in eq.~\eqref{eqn:NLGC_integral} can be computed analytically~\citep{2004ApJ...604..675S,2004JGRA..109.4107Z}. Comparing the parallel mean-free path $\lambda_{\parallel} = 3 \kappa_{\parallel} / v$ to the correlation length $\ell_{\text{c}}$, two limiting cases are noteworthy: For $\lambda_{\parallel} \ll \ell_{\text{c}}$ and for $\lambda_{\parallel} \gg \ell_{\text{c}}$, the results for $\lambda_{\perp}$ from QLT and from the nonlinear closure approximation of \citet{1974ApJ...191..235O} are recovered respectively. Note, however, that even though no assumption is made about the transport in the perpendicular directions (since $k_\perp = 0$ in slab turbulence) perpendicular transport turns out to be diffusive, again at variance with numerical test particle simulations (see Sec.~\ref{sec:applications_transport_coefficients}). For a composite slab+2D model, however, the NLGC theory agrees well with simulations if $a = \sqrt{1/3}$.

\subsubsection{Weakly non-linear theory}

In weakly non-linear theory (WLNT, \citealt{2004ApJ...616..617S}), the first two steps of NLGC theory are followed: (1) the factorisation of the fourth-order correlation function of two velocities and two magnetic field factors into two separate second-order correlation functions for velocities and magnetic field strength; (2) the decomposition of the field strength correlation function into the magnetic power spectrum and a characteristic function. The crucial difference with respect to the BAM theory is the form of the velocity correlations. Instead of eqs.~\eqref{eqn:BAM_Vxx} to \eqref{eqn:BAM_Vzz}, the QLT velocity correlations are kept for the perpendicular motions and only the parallel velocities are assumed to decorrelate at a rate $\omega$,
\begin{align}
V_{xx}(t) = V_{yy}(t) &= v^2 (1-\mu^2) \cos \Omega t \, , \label{eqn:WLNT_Vxx}
\\ -V_{xy}(t) = V_{yx}(t) &= v^2 (1-\mu^2) \sin \Omega t \, , \label{eqn:WLNT_Vyy}
\\ V_{zz}(t) &= v^2 \mu^2 \exp [- \omega t ] \, , \label{eqn:WLNT_Vzz}
\end{align}
where $\omega$ is identified with the pitch-angle scattering frequency, $\omega = 2 D_{\mu\mu} / (1 - \mu^2)$. For the characteristic function, a Gaussian distribution is assumed in the perpendicular direction whereas for the parallel motion, any possible diffuse contribution is ignored altogether.

Comparing the resulting expression with those from QLT it appears that only additional exponential factors with a linear time-dependence in the exponent have been introduced. When performing the time-integration these lead to resonance broadening which can be ascribed to pitch-angle scattering and perpendicular motion and the deviation of the particle orbits from purely helical motion. The resonance function is of the Breit-Wigner form. From this, the Fokker-Planck coefficient can be computed, in particular the pitch-angle diffusion coefficient and the perpendicular diffusion coefficient. Note however, that the perpendicular diffusion coefficient depends on the pitch-angle diffusion rate (or equivalently on the parallel diffusion coefficient). In order to probe the perpendicular diffusion independently when comparing to simulations, oftentimes the empirical parallel mean free path from the simulations is adopted.

\subsubsection{Other approaches}

\citet{2008ApJ...685L.165T} use a broadening of the resonance condition in isotropic turbulence, parametrised by smoothing of the particle position along the magnetic field as motivated by second-order QLT~\citep{2005PhPl...12e2905S}. The width of the particle position is computed from the usual QLT. As a consequence, $D_{\mu\mu}$ now has its maximum at $\mu=0$. The authors find good agreement with the numerical simulations of~\citet{Giacalone:1999}. Also noteworthy is the work of \citet{2009A&A...507..589S} who present an analytical computation of pitch-angle diffusion coefficient and mean-free path for slab turbulence. It is shown that QLT is a good approximation for $|\mu| >  \delta B / B_z$.

\section{Generating turbulent magnetic fields on a computer}
\label{sec:generating}

The most realistic way of generating a turbulent magnetic field on a computer to propagate particles in is of course to rely on simulations of this turbulence. This offers the opportunity to include (some of) the known complexity beyond the simple turbulence models described above, for instance anisotropic turbulence like the Goldreich-Sridhar picture~\citep{1994ApJ...432..612S,1995ApJ...438..763G}. Given the large dynamical range required for most applications, it is however also the most computationally expensive. In the following, we will review such attempts and their results, before discussing the generation of synthetic turbulence.

\subsection{Simulated turbulence}

The most extensive set of simulations to date have been performed by \citet{Cohet:2016goi}, CM16 from hereon, who tracked test particles through MHD turbulence generated with the \texttt{RAMSES} code~\citep{2002A&A...385..337T}. They followed the pioneering work of \citet{Beresnyak:2010yq} and \citet{Xu:2013ppa} and discussed differences in setups and results.

For the most part, CM16 ran the MHD part of their simulations on a $512^3$ grid, and the box length of the simulation was taken to be five times larger than the turbulence injection scale $L_{\text{inj}}$. This resulted in about one-and-a-half orders of magnitude in dynamical range between the coherence length of turbulence and the dissipation length, the latter being due to the finite numerical resolution. Turbulence was injected either by solenoidal or compressible forcing and the results differ significantly. It is hypothesised that this is  due to the preferential driving of Alfv\'enic turbulence for the solenoidal and of fast-magnetosonic turbulence for the compressible case, the latter leading to an isotropic turbulence cascade and being more efficient in CR scattering~\citep{Chandran2000,YanLazarian2002}.

CM16 studied in detail the dependence of parallel and perpendicular mean-free paths on the Alfv\'enic Mach number $M_{\text{a}}$ (which is defined as the ratio of the rms fluid velocity and the Alfv\'en speed in the total magnetic field, i.e.\ background plus turbulent). For the parallel mean-free path, a power law scaling with the Alfv\'enic Mach number $\lambda_{\parallel} \propto (M_{\text{A}})^{\alpha}$ is found. At small $M_{\text{a}}$, the results differ strongly between solenoidal and compressible forcing, with the parallel mean-free path at $M_{\text{a}} = 0.3$ being about two orders of magnitude larger in the former case. For the solenoidal case, $\lambda_{\parallel}$ is much larger than found by \citet{Xu:2013ppa} and the dependence on $M_{\text{A}}$ is much stronger: Typically $\alpha$ is between $-7$ and $-5$ which is also in tension with expectations from QLT where $\lambda_{\parallel} \propto M_{\text{A}}^{-2}$, e.g.\ \citep{2011ICRC...10..240S}. Note that this scaling was also confirmed in test particle simulations of synthetic isotropic turbulence, notably beyond the limits of validity of QLT~\citep{Casse:2001be}. For the compressible driving, $\lambda_{\parallel} \propto (M_{\text{A}})^{-2}$ as expected. The perpendicular mean-free path, on the other hand, is scaling like $\lambda_{\perp} \propto M_{\text{A}}^2$ in QLT which is largely confirmed by CM16. This is being ascribed to the contribution from field-line random walk to the perpendicular transport. Another prediction for compressible MHD turbulence~\citep{2008ApJ...673..942Y} is $\lambda_{\perp} \propto M_{\text{A}}^4$, but this only applies for the limits $\lambda_{\parallel} \ll L_{\text{inj}}$ or $\lambda_{\parallel} \gg L_{\text{inj}}$, whereas the simulations of CM16 are in between.

An equally crucial result is the dependence of the parallel and perpendicular mean-free paths on gyroradius $r_{\text{g}}$ (normalised with respect to the simulation scale $L$). Here, the results for $\lambda_{\parallel}$ again depend very sensitively on the driving at $L_{\text{inj}}$: If the forcing is solenoidal, the rigidity dependence of $\lambda_{\parallel}$ can be very weak: The dependence is power law like in the range of rigidities tested, $\lambda_{\parallel} \propto r_{\text{g}}^{\delta}$, and $\delta$ can become even negative, especially for large $M_{\text{a}}$. In QLT this is only possible for turbulence spectra $g(k) \propto k^{-q}$ with $q > 2$ while the power spectral indices found by CM16 are $q \sim 1.5$, that is consistently smaller than $2$. In the compressible case, the agreement with expectations is much better and the observed scaling is compatible with both $\delta = 1/3$ and $1/2$. (The dynamical range is too small to tell, in fact.) Perpendicular mean-free paths show less of a difference between the solenoidal and compressible cases and are largely consistent with a scaling $\propto r_{\text{g}}^{1/2}$. For gyroradii larger than $L_{\text{inj}}$, the transition to small-angle scattering with $\lambda_{\parallel} \propto r_{\text{g}}^2$ is being observed, as expected.

\subsection{Synthetic turbulence}

Realistic modelling of CR transport requires a rather wide dynamical range for the turbulent modes. MHD simulations of turbulence usually cover no more than one and a half orders of magnitude between the coherence length and the dissipation scale (see e.g.~CM16). An alternative to using simulated turbulence is to adopt one of the turbulence correlation tensors $P_{ij}(\vec{k}, t)$ discussed in Sec.~\ref{sec:turbulence_models} and to directly generate random realisations of a field with such a correlation structure on a computer. The turbulence generated in this way is usually referred to as ``synthetic turbulence''. The obvious drawback of this method is its reliance on a turbulence model instead of using the more realistic results from MHD simulations of turbulence. The advantages are the large dynamical range possible in principle, and the possibility of directly testing some of the  results of QLT and its non-linear extensions which are more straight-forward to compute for simple turbulence models.

When solving the equations of motion, we will need to evaluate the turbulent magnetic field $\vec{\delta B}$ at many different positions, possibly also at different times, the latter distinction becoming relevant when considering models of dynamical turbulence. In order to do this, we need to keep track not only of the amplitudes of the turbulent field, but also its phases which are random. This implies generating a random sequence of phases and storing them for the duration of the test particle simulation. On a computer, the turbulent magnetic field will be characterised by a finite number of real numbers, that is the corresponding magnetic field is band-limited. In the literature, two methods have been suggested, depending on whether the phases of a finite number of modes are stored or whether the turbulent magnetic field $\vec{\delta B}(\vec{r})$ is stored on a discrete grid. We will refer to the former as the \textit{harmonic method} and to the latter as the \textit{grid method}s. Both methods have their advantages, but also disadvantages which we will discuss.

\subsubsection{Harmonic method}

In the harmonic method, pioneered by~\citet{1994ApJ...430L.137G} and others~\citep{1997A&A...326..793M,1998A&A...337..558M,Giacalone:1999}, the turbulent field is defined as a superposition of plane waves,
\begin{equation}
\vec{\delta B}(\vec{r}) = \operatorname{Re} \left( \sum_{n=0}^{N-1} \delta \tilde{\vec{B}}_n \ee^{\ii \vec{k}_n \cdot \vec{r}} \right) \, ,
\end{equation}
Here, only the wavenumbers are discrete, and in order to cover as broad a dynamical range with as small a number $N$ of modes as possible, the spacing in $\vec{k}$ is oftentimes assumed to be logarithmic.

The alternative, but equivalent representation, 
\begin{equation}
\vec{\delta B}(\vec{r}) = \sum_{n=0}^{N-1} A_n \vhat{\xi}_n \cos \left[ k_n \vhat{k}_n \cdot \vec{r} + \beta_n \right] \, ,
\label{eqn:deltaB}
\end{equation}
makes explicit the interpretation as a superposition of $N$ independent waves travelling in the directions $\vhat{k}_n$ with amplitudes $A_n$, polarisations $\vhat{\xi}_n$, wavenumbers $k_n$ and phase factors $\beta_n$. Each mode $n$ is thus specified by six real numbers: one for $A_n$, one for $\vhat{\xi}_n$ (as it needs to be $\perp \vhat{k}_n$ in order for $\vec{\delta B}$ to be divergence-free), one for $k_n$, one for $\beta_n$ and two for $\vhat{k}_n$. Of these, $\vhat{\xi}_n$, $\vhat{k}_n$ and $\beta_n$ are random variables and their statistical distributions are determined by the turbulence model.

For instance, in isotropic turbulence (see Sec.~\ref{sec:3D_isotropic_turbulence}), $\vhat{\xi}_n$ is uniformly distributed on the unit circle (such that $\vhat{\xi}_n \cdot \vhat{k}_n = 0$), $\vhat{k}_n$ is uniformly distributed on the unit sphere and $\beta_n$ is uniformly distributed in $[0, 2 \pi [$. \cite{Giacalone:1999} suggested the following construction
\begin{equation}
\vec{\delta B}(x, y, z) = \sum_{n=1}^{N_m} A(k_n) \vhat{\xi}_n \exp \left[ \ii (k_n' z' + \beta_n) \right] \, , \label{eqn:sum_GJ1999}
\end{equation}
with polarisation vector
\begin{align}
\vhat{\xi}_n &= \cos \alpha \, \vhat{x}_n' + \ii \sin \alpha \, \vhat{y}_n' \label{eqn:polarisation_GJ1999} \, ,
\end{align}
and
\begin{equation}
\arraycolsep=2.7pt
\left( \!\! \begin{array}{c} x' \\ y' \\ z' \end{array} \!\! \right)
\!\! = \!\!
\left( \!\! \begin{array}{c c c} 
\cos \theta_n \cos \phi_n 	& \cos \theta_n \sin \phi_n 	& - \sin \theta_n \\ 
- \sin \phi_n 			& \cos \phi_n 				& 0 \\ 
\sin \theta_n \cos \phi_n 	& \sin \theta_n \sin \phi_n 		& \cos \theta_n 
\end{array} \!\! \right) \!\!\!
\left( \!\! \begin{array}{c} x \\ y \\ z \end{array} \!\! \right)
\label{eqn:rotation_GJ1999}
\end{equation}

These equations describe a superposition of waves with wavenumbers $k_n$ and (complex) amplitudes $A(k_n)$. The direction of each mode is along the $z'$-axis in a coordinate system generated from the lab system through a rotation by $\theta_n$ around the $y$-axis and a subsequent rotation by $\phi_n$ around the new $z'$-axis. $\{ \theta_n, \phi_n, 0 \}$ are thus the Euler angles defining the rotation of the lab system into the rotated system in the $zyz$ convention. Note that the first term in the exponent of eq.~\eqref{eqn:sum_GJ1999} has been simplified in primed coordinates, $\vec{k} \cdot \vec{x} = \vec{k}' \cdot \vec{x}' = k_z' z'$.

It has been claimed~\citep{Tautz:2013jd} that this construction does not guarantee the correct variances for all three components of the turbulent magnetic field. We believe, however, that \citet{Tautz:2013jd} did not compute the averages correctly and that with appropriate averages, the construction by Giacalone and Jokipii give the correct results.

The $A_n$ in turn are fully determined by the power spectrum of turbulence. Again, for an isotropic turbulence tensor, $\avg{ \delta\tilde{B}_i(\vec{k}) \delta\tilde{B}_j(\vec{k}') } = \delta_{ij} \delta^{(3)} (\vec{k} - \vec{k}') g(k)$ and thus $A_n = \sqrt{g(k_n)}$ is the discrete approximation for the desired power spectrum.

While the turbulence model fixes the $A_n$ and the statistical distributions of the $\vhat{\xi}_n$, $\vhat{k}_n$ and $\beta_n$, what is not fixed is the binning of the $k_n$ and the total number of modes, $N$. Both are usually constrained by the need to cover as wide a dynamical range as possible. Given our understanding from QLT that interactions are resonant, what is required in the magneto-static limit for one particle energy at a minimum is a spectrum spanning at least a factor of a few around the resonant wavenumber. In addition, power on larger scales can have an impact, depending exactly on what the observable is. This means that easily a few orders of magnitude in wavenumber range are required, even at minimum. Therefore, oftentimes a logarithmic spacing in $k$ is adopted. This leaves open the question what the required number $N$ of modes is. For the case of slab-turbulence, this question has been investigated using the convergence with number of modes of a ``quasi-Lyapunov exponent''~\cite{Tautz:2013jd}. On a more practical level, we note that the number oftentimes adopted are $N = \mathcal{O}(100)-\mathcal{O}(1000)$ for a dynamical range $k_{\text{min}} / k_{\text{max}} \sim 10^4$.

\subsubsection{Grid method}

\begin{figure*}[!th]
\includegraphics[width=\textwidth]{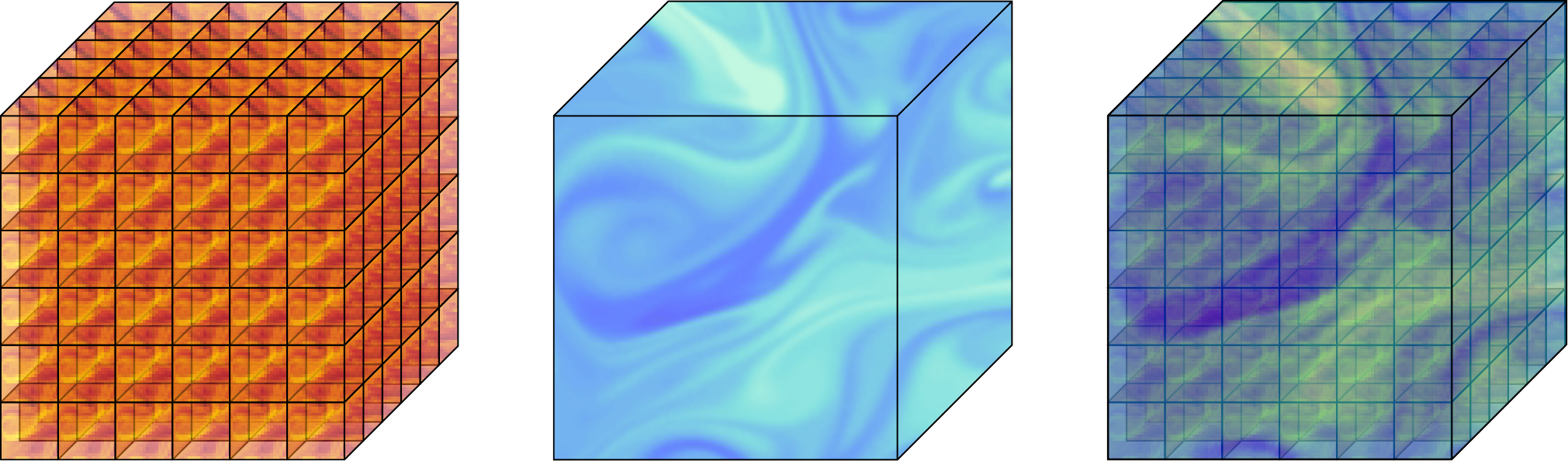}
\caption{Illustration of the idea of using nested grids. Note that in this illustration padding is not used and thus the grids are not overlapping.}
\label{fig:nested_3D}
\end{figure*}

\paragraph{Standard grid method.} An alternative way to set up turbulent magnetic fields on a computer is called the grid method (e.g.~\citealt{2002GeoRL..29.1048Q}). While in the harmonic method the amplitudes and phases of the turbulent modes are stored (e.g. in the combination $\{ A_n, \vhat{\xi}_n, \vhat{k}_n, \phi_n \}$, in the grid method the turbulent magnetic field itself $\vec{\delta B}(\vec{r})$ is stored on a spatial grid $\vec{r}_{i,j,k}$ and can be interpolated between these grid points.

Here, we introduce the discretisations of the position $\vec{r}_{n_1,n_2,n_3} = (x_{n_1}, y_{n_2}, z_{n_3})^T = (n_1 \Delta r_1, n_2 \Delta r_2, n_3 \Delta r_3)^T$ and wavenumber $\vec{k}_{m_1,m_2,m_3} = (k_x)_{m_1}, (k_y)_{m_2}, (k_z)_{m_3})^T \break= (m_1 \Delta k_1, m_2 \Delta k_2, m_3 \Delta k_3)^T$. The Fourier transform pair of eqs.~\eqref{eqn:FT_pair_1} and \eqref{eqn:FT_pair_2}, $\delta \tilde{B}_j(\vec{k})$ and $\delta B_j(\vec{x})$, then corresponds to the discrete Fourier transform pair $\delta \tilde{B}_j^{m_1, m_2, m_3}$ and $\delta B_j^{n_1, n_2, n_3}$,
\begin{widetext}
\begin{align}
\delta \tilde{B}_j^{m_1, m_2, m_3} &= \sum_{n_1=0}^{N_1-1} \sum_{n_2=0}^{N_2-1} \sum_{n_3=0}^{N_3-1} \ee^{2 \pi \ii (\frac{m_1 n_1}{N_1} + \frac{m_2 n_2}{N_2} + \frac{m_3 n_3}{N_3})} \delta B_j^{n_1, n_2, n_3} \, , \\
\delta B_j^{n_1, n_2, n_3} &= \frac{1}{N_1 N_2 N_3} \sum_{m_1=0}^{N_1-1} \sum_{m_2=0}^{N_2-1} \sum_{m_3=0}^{N_3-1} \ee^{-2 \pi \ii (\frac{m_1 n_1}{N_1} + \frac{m_2 n_2}{N_2} + \frac{m_3 n_3}{N_3})} \delta \tilde{B}_j^{m_1, m_2, m_3} \, ,
\end{align}
\end{widetext}

\noindent for discretely sampled $\delta B_i(\vec{r})$ and $\delta \tilde{B}_i(\vec{k})$,
\begin{align}
\delta \tilde{B}_j^{m_1, m_2, m_3} &= \frac{(2 \pi)^{3/2}}{\Delta x_1 \Delta x_2 \Delta x_3} \delta \tilde{B}_j(\vec{k}_{m_1,m_2,m_3}) \, , \\
\delta B_j^{n_1, n_2, n_3} &= \delta B_j(\vec{r}_{n_1,n_2,n_3}) \, .
\end{align}%

A fast way of setting up a homogeneous scalar Gaussian random field in 3 dimensions with a given power spectrum works in harmonic space. The requirement
\begin{equation}
\avg{ \delta \tilde{B}_i(\vec{k}) \delta \tilde{B}^*_j(\vec{k}) } = P_{ij}(\vec{k}) \, ,
\end{equation}%
only fixes the amplitudes, but not the complex phases. To obtain a homogeneous Gaussian random field (with the correlation structure defined by the power spectrum), the phases must be complex normal distributed,
$\arg(\delta \tilde{B}_n) \sim \mathcal{N}(0,1) + \ii \, \mathcal{N}(0,1)$. However, for a \emph{real} turbulent field the phases need to further satisfy the relation implied by eq.~\eqref{eqn:reality}. For a discrete field in one dimension, that is $\delta \tilde{B} (N/2-k_n) = \delta \tilde{B}^* (k_n)$. Instead of enforcing the reality conditions by hand, it has proven convenient to use an efficient routine for the generation of a real Gaussian random field with no correlation structure, that is white noise, Fourier transform and then scale the complex amplitudes with the desired power spectrum before transforming back. Note that modern Fourier transform libraries provide routines for reconstructing the full inverse Fourier transform from the Fourier transform at just the positive (spatial) frequencies.

Knowing how to generate a scalar Gaussian random field, it might seem that we just need to combine three independent scalar fields into a 3D vector. However, in general this 3D random field will not be divergence-free. In order to guarantee that the field is divergence-free, only the polarisations perpendicular to $\vhat{k}$ should be retained. This can be achieved by subtracting from each $\delta \tilde{B}_j^{m_1, m_2, m_3}$ the projection of it onto $\vhat{k}$.

The advantage of the grid method is most importantly its speed: Instead of performing a sum of $N$ modes for a large number of test particles at each timestep of the test particle propagation, only an interpolation between the relevant grid points is needed. For a fine enough grid in 3D, a tri-linear interpolation is sufficient. (See, however, \citet{Schlegel:2019uww}.) In most cases, this is computationally more efficient. However, this gain in speed is achieved at the price of increased memory requirements. For example, a 3D field of doubles on a $2048^3$ grid requires $192 \, \text{GB}$ of RAM, where we have ignored overhead. While certain nodes of computing clusters can have more RAM, as of the writing of this \textit{review}, this is already beyond the reach but of the most powerful personal computers.

At any rate, a finite grid size implies issues with periodicity and accuracy of interpolation. The latter can be minimised by ensuring that the smallest wavenumber are a factor of a few larger than the grid spacing, $\lambda_{\text{min}} = (\text{a few}) \, \Delta x$. At the same time, a few of the largest modes should fit onto the extent $L$ of the grid, $L = (\text{a few}) \, \lambda_{\text{max}}$, in order to reduce possible periodicity issues. Thus with 2048 grid points, we can cover at most a dynamical range of $\lambda_{\text{max}} / \lambda_{\text{min}} \sim \mathcal{O}(100)$. This is probably enough to capture the particle-wave resonance, even for broadened resonances. However, modes at scales larger than the resonant scale can also have an effect on particle transport, e.g.\ through FLRW, but cannot be taken taken into account for such a small dynamical range.

\begin{figure*}[thb]
\includegraphics[width=\textwidth]{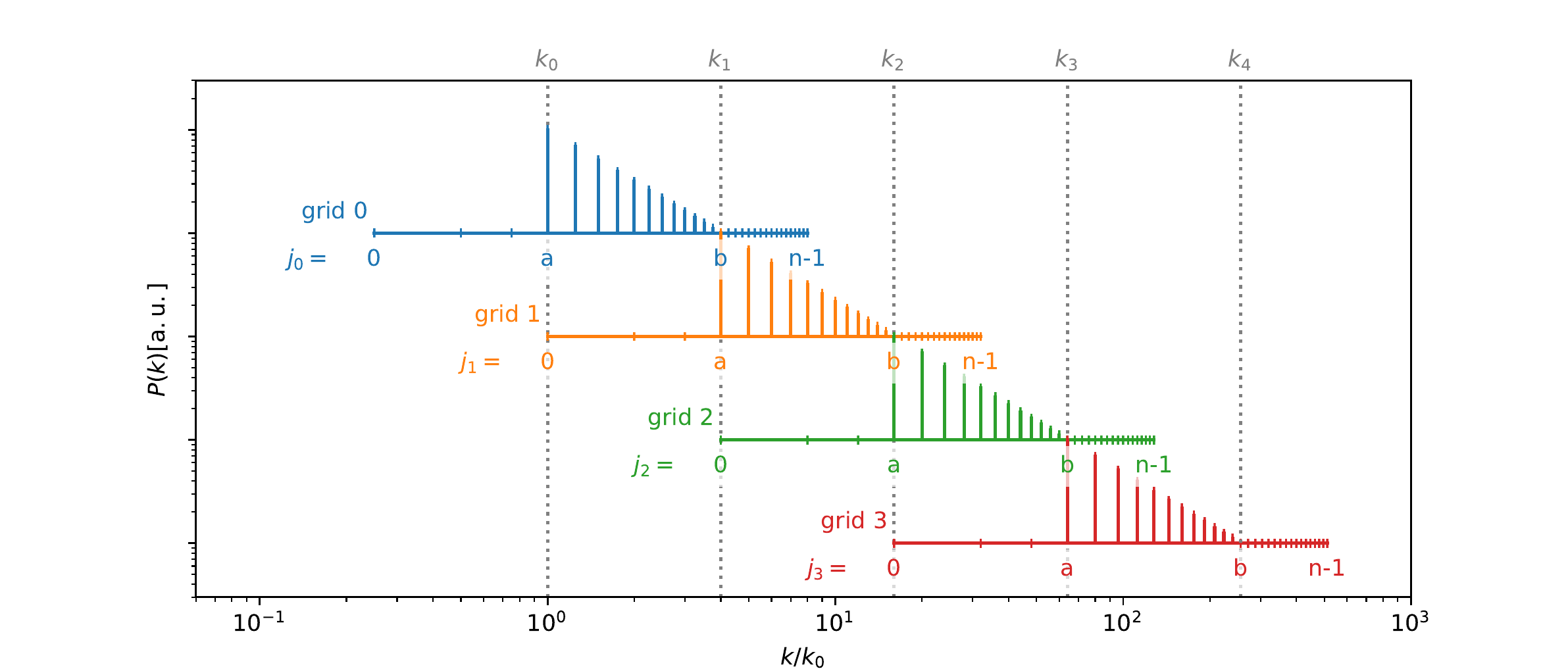}
\caption{Illustration of the nested grid approach. Shown is the power spectrum and how it is partitioned onto four sub-grids $i$, each only contributing in a limit range of wavenumbers.}
\label{fig:nested_sketch}
\end{figure*}

\paragraph{Nested grid method.} In light of these considerations, it was suggested~\citep{2012JCAP...07..031G} to increase the dynamical range by using nested grids. This method was later also used by \citet{Mertsch:2014cua} and \citet{2015ApJ...809L..23S}. The idea is that the total dynamical range $[ k_{\text{min}}, k_{\text{max}}]$ is divided into $N$ intervals $[ k_i, k_{i+1}]$ with $k_0 =k_{\text{min}}$ and $k_{N+1} = k_{\text{max}}$. Each interval is set up on a separate grid and these sub-grids are then periodically replicated over the whole computational domain. See Fig.~\ref{fig:nested_3D} for an illustration of the method in 3D.

The total turbulent field is given by the sum of turbulent fields on each grid. For a power law power spectrum, $P(k) \propto k^{-q}$, the turbulent energy $\delta B_i^2$ to be localised on a sub-grid $i$ is
\begin{equation}
\delta B_i^2 = \delta B^2 \frac{k_i^{3-q} - k_{i+1}^{3-q}}{k_{\text{min}}^{3-q} - k_{\text{max}}^{3-q}} \, .
\end{equation}

As for the case of a single grid, it is advisable not to use the whole range of the grid for turbulent modes, but to use part of the range for padding. In Fig.~\ref{fig:nested_sketch}, we illustrate the overlapping nested grids produced by this construction.

In this way a much larger dynamical range can be achieved. For definiteness, we close the discussion of nested grids with an example for how to set up the (sub-)grids for a test particle simulation. In Fig.~\ref{fig:nested_sketch}, we illustrate the nesting of four grids with 32 points each. On each grid $i$, we are only using 12 points to set up the turbulent modes with a dynamical range of $k_{i+1}/k_i = 12$. The remaining 20 points are used for padding. For example, we can set the amplitude to zero for the first $(a-1) = 3$ modes, have finite power between $j_i = a$ and $b$ (corresponding to the wavenumbers $k_i$ and $k_{i+1}$) and again no power for the remaining grid points. Note how the wavenumber grids are organised in order for the different grids to smoothly connect.

The parameters of this examples have been chosen to allow for a clear presentation in Fig.~\ref{fig:nested_sketch}. As a real application example, we might instead consider the propagation of $10 \, \text{TeV}$ test particles in a $\sqrt{\avg{B^2}} = 4 \, \mu\text{G}$ isotropic field with a $k_{\text{min}}$ of $0.1 \, \text{kpc}$. The gyroradius in the $4 \, \mu\text{G}$ field is $\sim 2.7 \times 10^{-6} \, \text{kpc}$, thus the dynamical range required is at least $0.1 / (2.7 \times 10^{-6}) \simeq 3.7 \times 10^4$. This could be achieved by nesting five grids of 128 points each, each grid only covering a factor $16$ in dynamical range. The remaining range of $128/16 = 8$ would be used for padding. Note that without nesting, the dynamical range of $3.7 \times 10^4$ would have required a number of grid points per dimension of $131\,072$ or more which corresponds to $48 \, \text{PB} $ of RAM for a 3D-vector field of doubles!

\section{Applications}
\label{sec:applications}

Traditionally, test particle simulations have been used primarily for computation of diffusion coefficients which would then be compared with analytical results in order to test CR transport theories~\citep{Giacalone:1999,2007JCAP...06..027D,Snodin:2015fza,2017ApJ...837..140S}. In addition, test particle simulations have been used (and are still being used) to study the deflection of ultra-high energy CRs in the Galactic magnetic fields where transport is certainly not resonant pitch-angle scattering~\citep{1971ICRC....1..310K,Harari:2000az,Tinyakov:2001ir,AlvarezMuniz:2001vf,Harari:2002dy,Kachelriess:2005qm,2014APh....54..110B,2019JCAP...05..004F}. There are however situations where even Galactic transport is not diffusive or where the diffusive picture is questionable. These include the escape of Galactic CRs from the CR halo around the knee~\citep{2007JCAP...06..027D,Giacinti:2015hva}, near source transport~\citep{Giacinti:2012ar,Kachelriess:2015oua}, stochastic acceleration~\citep{OSullivan:2009rvg,Winchen:2016koj}, also in relativistic turbulence~\citep{Demidem:2019jzn}, and the study of CR anisotropies~\citep{Giacinti:2011mz,Giacinti:2011ww,Schwadron2014,Mertsch:2014cua,Ahlers:2015dwa,Lopez-Barquero:2015qpa,Pohl:2015fdp,Kumar:2018qwa,Mertsch:2019xij}. In the following, we will briefly review the use of test particle simulations and discuss the results for a few physics cases.

\subsection{Computing transport coefficients}
\label{sec:applications_transport_coefficients}

All the non-linear extensions that are meant to address QLT's issues need to make certain assumptions (see Sec.~\ref{sec:extensions}). While these assumptions may be well motivated, it is not clear \textit{a priori} whether they result in an accurate description of CR transport. It is therefore of great interest to test these theories by comparing their results with those of numerical simulations.

A central prediction of the non-linear models are the parallel and perpendicular mean-free path or equivalently the parallel and perpendicular diffusion coefficients, $\kappa_{\parallel}$ and $\kappa_{\perp}$. To a lesser extent, numerical simulations have also been employed to compute the pitch-angle scattering diffusion coefficients $D_{\mu\mu}$ and the off-diagonal, anti-symmetric elements of the diffusion tensor $\kappa_{\text{A}}$ describing drifts. Of course, checking if transport is diffusive in the first place (instead of subdiffusive or superdiffusive) is another important application of test particle simulations.

\subsubsection{Technical details}

We start by recalling the definition of the instantaneous diffusion coefficients,
\begin{align}
d_{ii}(t) = \frac{\avg{ (\Delta x_i)^2 }}{2 t} \, .
\end{align}
The mean square displacements $\avg{ (\Delta x_i)^2 }$ are directly accessible for a set of trajectories $\{ \vec{r}_j \}$ from test particle simulations
\begin{equation}
\avg{ (\Delta x_i)^2 } = \avg{ |r_{j,i}(t) - r_{j,i}(0)|^2 } \, . \label{eqn:square_displacement_trajectories}
\end{equation}
Assuming again that the regular magnetic field $\avg{ \vec{B} } = \Bregular{} \vhat{z}$, we identify $d_{\parallel} = d_{zz}$ and $d_{\perp} = d_{xx} = d_{yy}$.

As far as the averaging on the RHS of eq.~\eqref{eqn:square_displacement_trajectories} is concerned, most authors have adopted an averaging over initial particle velocity and over magnetic field realisations. The former is necessary as the (instantaneous) diffusion coefficients do not retain any pitch-angle dependence, cf.\ eq.~\eqref{eqn:kappa_parallel}, and the latter is a consequence of QLT considering the ensemble-averaged phase space density. There is no agreement in the literature, however, on how many particle directions and how many field realisations are required to accurately compute diffusion coefficients. 

For times much larger than the scattering time, the instantaneous diffusion coefficients should converge towards the asymptotic diffusion coefficients, $\kappa_{\parallel}$ and $\kappa_{\perp}$. Depending on the normalised rigidity, that is the gyroradius divided by the correlation length, $r_{\text{g}}/l_{\text{c}}$, and on the level of turbulence, this only happens after many gyroperiods. Correspondingly, the computational expense can be very high. In order to increase the statistics at intermediate times, it was suggested~\citep{Casse:2001be} to not only use the initial position $\vec{r}_j(0)$ as one endpoint of simulated trajectories in computing the mean squared distances, but to also consider intermediate intervals $[t_i, t_{i+1}]$. This improves the statistics of trajectories for intermediate times, however, it is not clear whether this does not introduce some unwanted correlations.

We note that it is also possible to test the diffusion approximation by computing $\kappa_{\parallel}$ from the pitch-angle diffusion coefficient $D_{\mu\mu}$. Note that in practice, oftentimes the scattering rate is derived from the already pitch-angle averaged correlation function $\avg{ \mu(t) \mu(0) }$ instead of from the pitch-angle diffusion coefficient $D_{\mu\mu}(\mu)$.

We note that already \citet{Giacalone:1999} explored alternatives for computing the diffusion coefficients. The solution of the diffusion equation for an initially localised distribution is a multi-variate Gaussian with variances $\sigma_{\parallel} = 2 \kappa_{\parallel} t$ and $\sigma_{\perp} = 2 \kappa_{\perp} t$ in the parallel and perpendicular directions. Determining the spread of a set of trajectories from their common origin therefore allows computing the diffusion coefficients.

\subsubsection{Results}

In the following, we provide a brief overview of some of the first and some more recent computations of diffusion coefficients using test particle simulations. The first to use test particle simulations for the computation of transport coefficients were \citet{1994ApJ...430L.137G}. Considering simplified turbulence models with 2D and 3D magnetostatic, isotropic turbulence they showed that diffusive perpendicular transport required 3D turbulence. For that case, they numerically computed $\kappa_{\parallel}$ and $\kappa_{\perp}$ for the first time. The ratio ($\kappa_{\perp}/\kappa_{\parallel})$ was found to deviate from the prediction of classical scattering (see Sec.~\ref{sec:derivation_FPE}) which was ascribed to the small dynamical range of the turbulence spectrum. This pioneering paper was followed up on by Micha{\l}ek and Ostrowski a few years later~\citep{1997A&A...326..793M,1998A&A...337..558M}. Adopting the same harmonic method as Giacalone and Jokipii, Micha{\l}ek and Ostrowski already considered a more complex and realistic scenario, including time-dependent turbulence with electric fields, that allowed them to study the role of stochastic acceleration and compare both parallel spatial diffusion and momentum diffusion with the prediction of QLT~\citep{1996NPGeo...3...66M}. For slab-like turbulence they found good agreement. They also took into account the proper polarisation properties of the linear MHD waves in the cold plasma limit, that is shear-Alfv\'en and  fast magnetosonic waves~\citep{1998A&A...337..558M}. They found a much more effective cross-field diffusion for magnetosonic waves, compared to the case with Alfv\'en waves.

However, all these early simulations exhibited a rather limited dynamical range, $\lesssim 100$. A follow-up of their earlier work, \citet{Giacalone:1999} extended this dynamical range to $10^4$. A first study employing not only the harmonic method, but also the grid-based approach was performed by \citet{Casse:2001be} and they showed that both methods gave similar results. \citet{2002GeoRL..29.1048Q} shifted the focus back to perpendicular transport and showed that in close-to slab turbulence, perpendicular transport is indeed compound subdiffusion. This is the case as long as there is too little structure in the perpendicular directions. If however there is sufficient structure, a second regime of diffusion is attained after a transitory phase of subdiffusion~\citep{2002ApJ...578L.117Q}.

Another study with interest in applications to Galactic transport was the one by \citet{2007JCAP...06..027D}. Not only did they consider extended turbulence and rigidity ranges, but also the mixed effects of particle scattering and drifts due to small-scale turbulence and inhomogeneities in the large-scale regular, background field. This is to be expected for the Galactic magnetic fields which are thought to trace out a spiral structure, similar to gas and stars in the Galaxy. It was shown that both diffusion and drifts can play an important role in the escape of cosmic rays from the extended halo around $10^{17} \, \text{eV}$. \citet{2012JCAP...07..031G} followed a similar interest in the transport of Galactic cosmic rays at PeV energies and the possibility to use the predicted dipole anisotropies to set limits on the Galactic contribution at even higher energies. To simulate particles at rigidities of $30 \, \text{PV}$, they need to cover a dynamical range that was beyond the use of single grids for the turbulent magnetic field and they adopted nested grids instead. More recently, \citet{Giacinti:2017dgt} have also simulated anisotropic turbulence structures. \citet{Snodin:2015fza} have presented simulation results for the largest dynamical range yet. They have considered 3D isotropic turbulence with a broken power law spectrum, motivated by the need for heuristic description of diffusion coefficients for MHD Galaxy/ISM simulations. They also took into account the contribution from FLRW, fitting the field line diffusion coefficient from the simulated turbulence.

\begin{figure*}[]
\includegraphics[scale=1]{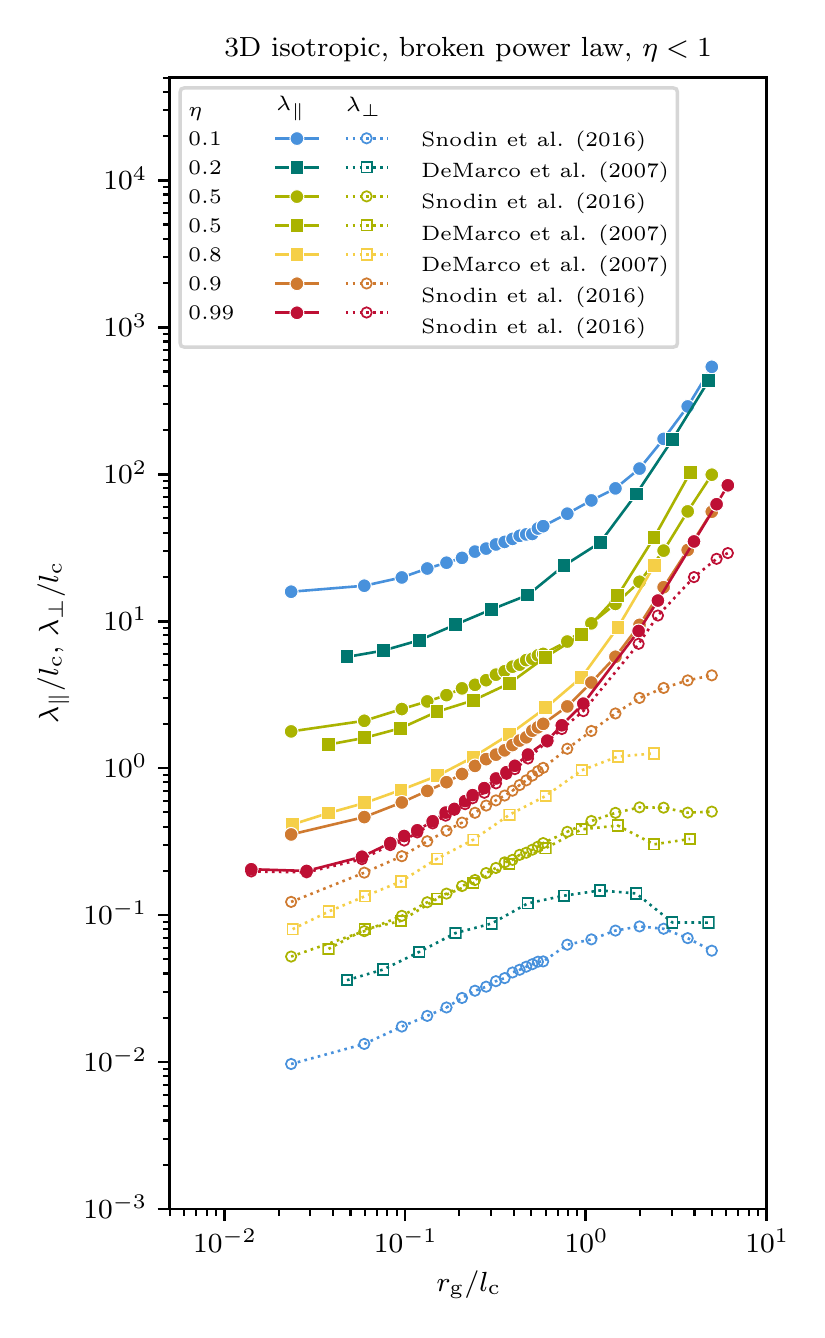}
\includegraphics[scale=1]{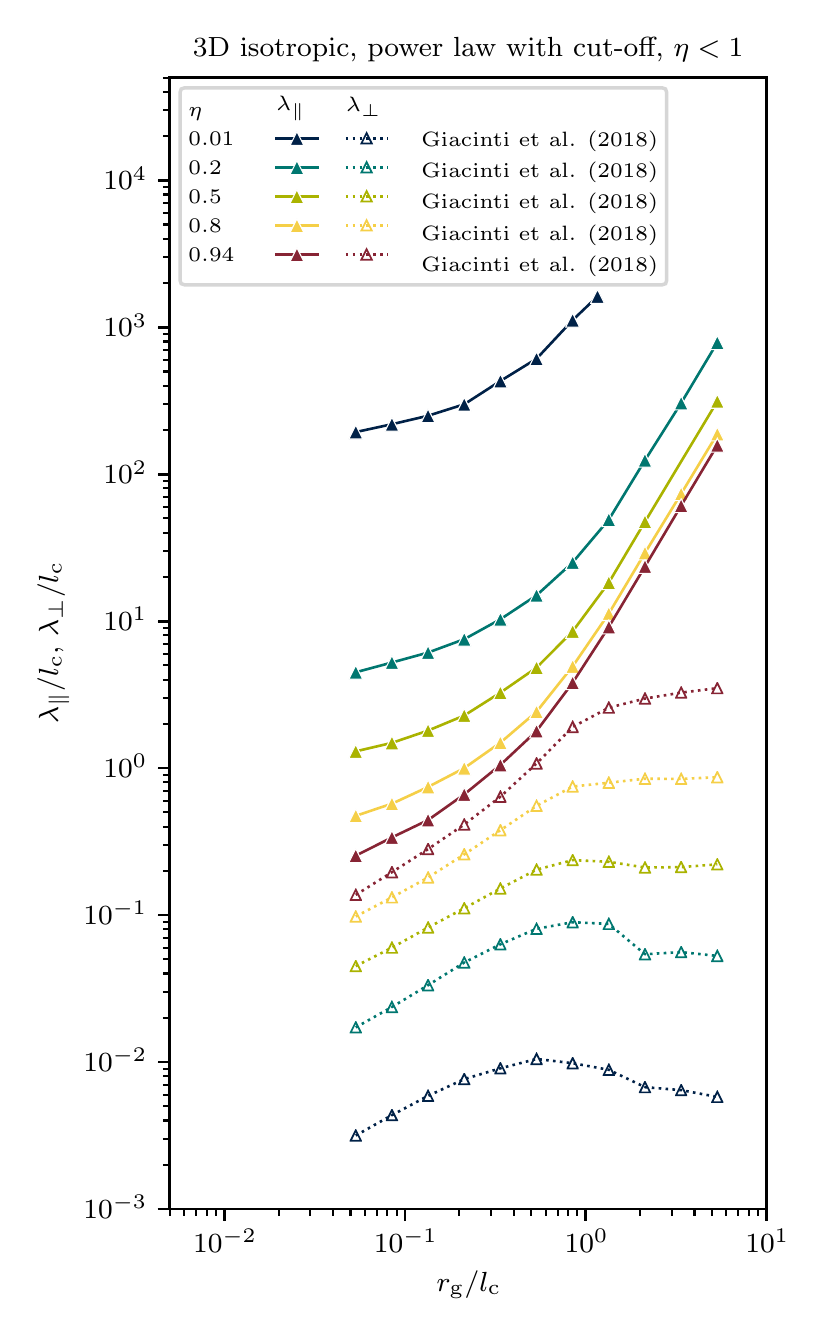} \\
\includegraphics[scale=1]{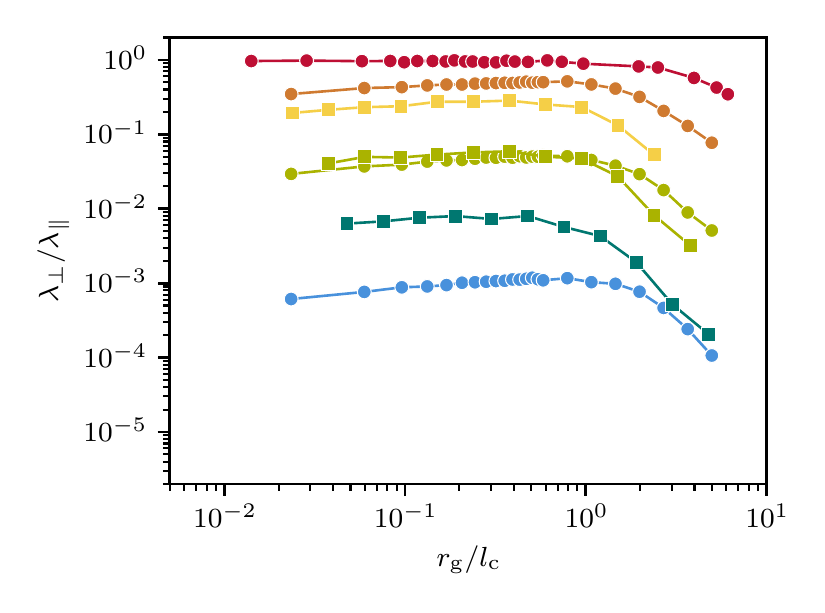}
\includegraphics[scale=1]{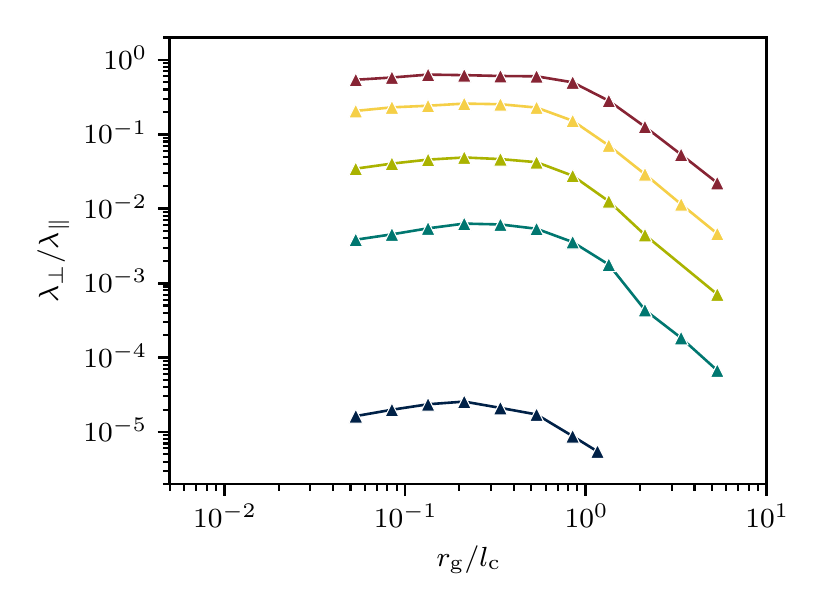} \\
\caption{Compilation of mean free paths (normalised to the correlation length $l_{\mathrm{c}}$), computed from test particle simulations as a function of gyroradius $r_{\text{g}}$ for isotropic turbulence (also normalised to $l_{\mathrm{c}}$). \textbf{Top left:} Parallel and perpendicular mean free paths $\lambda_{\parallel}$ and $\lambda_{\perp}$, assuming a broken power law turbulence spectrum, eq.~\eqref{eqn:BPL}, for various values of the turbulence level $\eta \equiv \delta B^2 / (B_0^2 + \delta B^2)$. \textbf{Top right:} $\lambda_{\parallel}$ and $\lambda_{\perp}$, assuming a power law turbulence spectrum with cut-off, eq.~\eqref{eqn:PL}. \textbf{Bottom left and right:} Ratio $\lambda_{\perp} / \lambda_{\parallel}$, again for a broken power law spectrum and a power law turbulence spectrum with cut-off, respectively.}
\label{fig:compilation}
\end{figure*}

\begin{figure*}[t]
\includegraphics[scale=1]{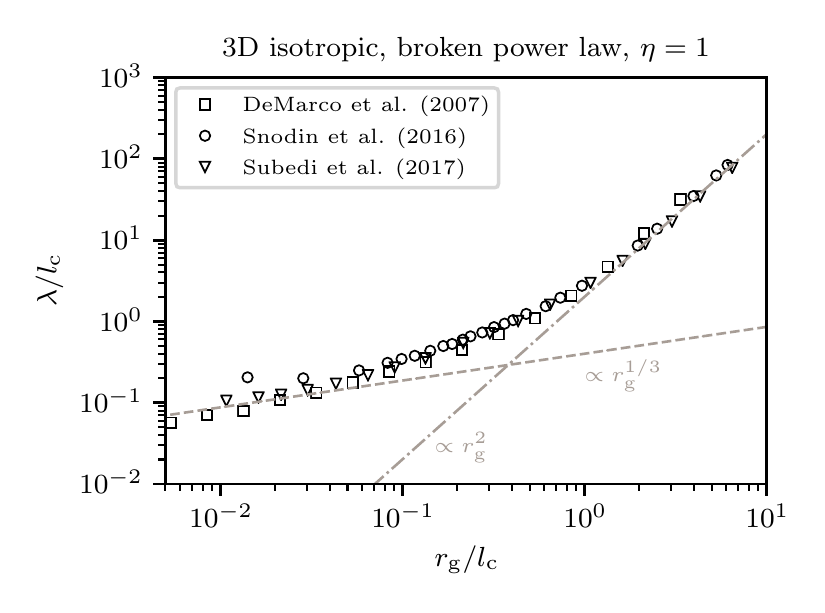}
\includegraphics[scale=1]{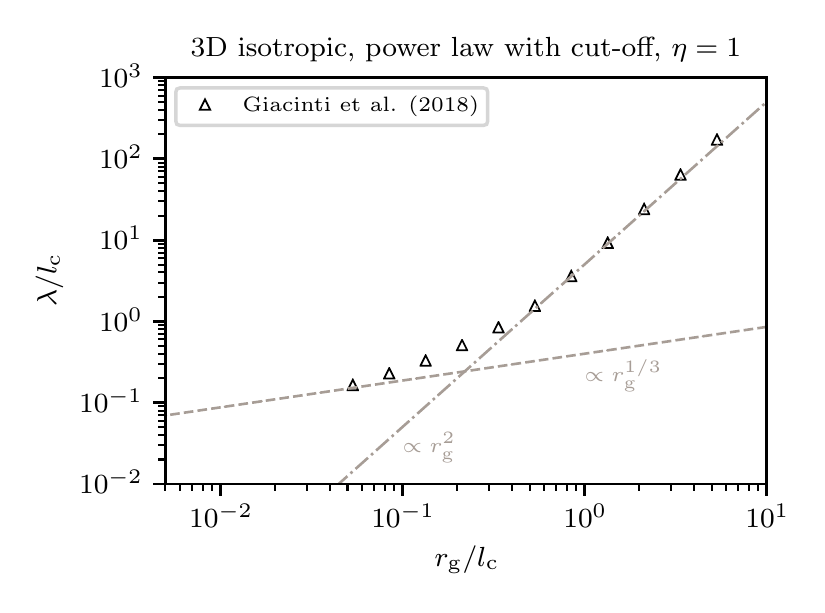}
\caption{Compilation of mean free paths (normalised to the correlation length $l_{\mathrm{c}}$), computed from test particle simulations as a function of gyroradius $r_{\text{g}}$ for isotropic turbulence (also normalised to $l_{\mathrm{c}}$). \textbf{Left:} Mean-free path $\lambda$, assuming a broken power law turbulence spectrum, eq.~\eqref{eqn:BPL}, without regular field, that is $\eta = 1$. \textbf{Right:} $\lambda$, assuming a power law turbulence spectrum with cut-off, eq.~\eqref{eqn:PL}, without regular field, that is $\eta = 1$.}
\label{fig:compilation2}
\end{figure*}

\begin{figure}[!t]
\includegraphics[scale=0.99]{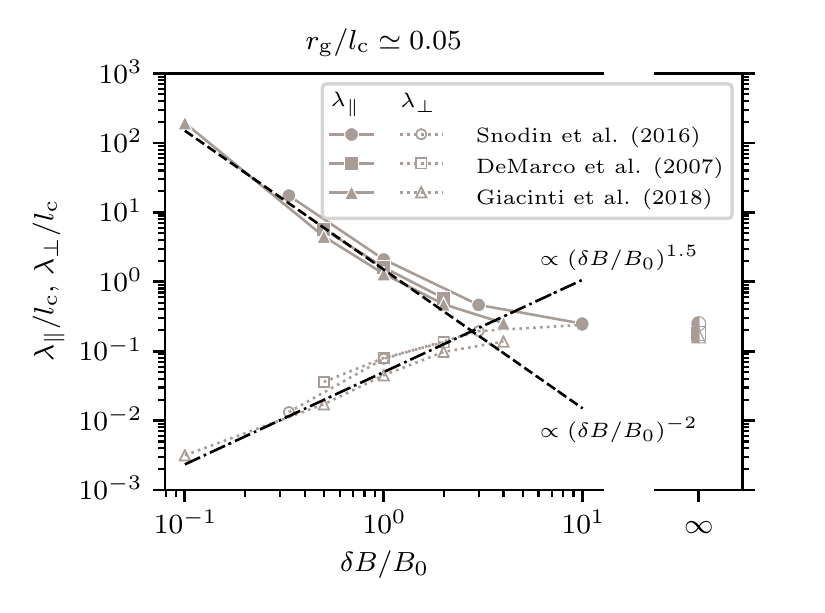}
\caption{Compilation of mean free paths (normalised to the correlation length $l_{\mathrm{c}}$), computed from test particle simulations as a function of $(\delta B/B_0)$. The dashed and dash-dotted line show power laws $\propto (\delta B/B_0)^{-2}$ and $\propto (\delta B/B_0)^{1.5}$, respectively.}
\label{fig:dB2B0}
\end{figure}

In Fig.~\ref{fig:compilation}, we present a compilation of mean free paths computed for a 3D isotropic turbulence model and two different spectral shapes: a broken power law, see eq.~\eqref{eqn:BPL} in the left panels, and a power law with cut-off, eq.~\eqref{eqn:PL} in the right panels. In the top panels, we show the parallel and perpendicular mean free paths $\lambda_{\parallel}$ and $\lambda_{\perp}$ as a function of the gyroradius $r_{\text{g}}$ for various levels of turbulence, $\eta \equiv \delta B^2 / (B_0^2 + \delta B^2)$. (The simulations have been performed at discrete energies, of course, and the points are connected only to guide the eye.) The different scalings of $\lambda_{\parallel}$ and $\lambda_{\perp}$ with $r_{\text{g}}$ in the regimes of resonant scattering ($r_{\text{g}} \ll l_{\text{c}}$) and small-angle scattering ($r_{\text{g}} \gg l_{\text{c}}$) are clearly visible. The behaviour with $\eta$ is largely montonic. There is agreement between different groups that have presented results for the same setup, for instance compare the light green lines ($\eta = 0.5$) in the top left panel. In the resonant scattering regime, the mean-free paths seem to agree even for the different spectral shapes, compare the light green lines in the top left and top right panels for $r_{\text{g}}/l_{\text{c}} \lesssim 1$. At larger gyroradii, $r_{\text{g}}/l_{\text{c}} \gg 1$, the agreement is worse, however, with the case with a power law with cut-off exhibiting somewhat larger parallel and smaller perpendicular mean-free paths. This behaviour was to be expected as for $r_{\text{g}}/l_{\text{c}} \gg 1$, the particle transport becomes more sensitive to the large modes of the turbulent spectrum where the difference between the spectra is most stark.

In the lower panels of Fig.~\ref{fig:compilation}, we show the ratio of perpendicular and parallel mean free paths, $\lambda_{\perp} / \lambda_{\parallel}$. The agreement between different groups is again fair. It appears that at low turbulence levels, $\lambda_{\perp}$ has a stronger dependence on $r_{\text{g}}$ than $\lambda_{\parallel}$, but this requires further tests. 

It becomes apparent from the top panels of Fig.~\ref{fig:compilation}, that the parallel and perpendicular mean free paths start converging towards the isotropic mean-free path $\lambda$ as $\eta \to 1$. (For clarity, we have plotted the limit $\eta = 1$, that is no background field, separately in Fig.~\ref{fig:compilation2}.) Taken together, the simulations stretch three orders of magnitude in gyroradius and the different scalings in the low- and high-rigidity regimes is easily identified as $\lambda \propto r_{\text{g}}^{1/3}$ for $r_{\text{g}} \ll l_{\text{c}}$ and $\lambda \propto r_{\text{g}}^2$ for $r_{\text{g}} \gg l_{\text{c}}$. The agreement between different groups for the same spectral shape is excellent and the results for different spectral shapes are most pronounced in the small-angle scattering regime where the mean-free path is again larger for the power law with cut-off again than for the broken-power law shape.

In Fig.~\ref{fig:dB2B0} we show the dependence of the parallel and perpendicular mean free paths on $(\delta B/B_0)$. While the parallel mean free paths are $\propto (\delta B/B_0)^{-2}$, as expected from QLT, the perpendicular mean free paths are closer to $\propto (\delta B/B_0)^{1.5}$ whereas QLT (without the FLRW contribution) predicts $\propto (\delta B/B_0)^2$. We emphasise again that for isotropic turbulence, QLT is not valid as it predicts an infinite parallel mean-free path.

In Tbl.~\ref{tbl1} we compare the prediction of $\kappa_{\parallel}$ and $\kappa_{\perp}$ from various transport theories to the results from numerical simulations. \\

\begin{table*}[!tbh]
\small
\caption{Comparison of parallel and perpendicular transport in simulations and theories for different turbulence geometries. Here, we assume magnetostatic turbulence.\label{tbl1}}
\begin{tabular}{l || l | l | l}
\tableline
						& isotropic 											& slab								& composite \\
\tableline\tableline
\multirow{2}{*}{simulations}	& $\parallel$ diffusive 									& $\parallel$ diffusive					& $\parallel$ diffusive \\
						& $\perp$ diffusive\tablenotemark{a}	& $\perp$ subdiffusive ($\kperp \propto 1/\sqrt{t}$)	& $\perp$ diffusive \\
\tableline\tableline
\multirow{2}{*}{QLT}			& $\parallel$ superdiffusive ($\kparr \to \infty$) \xmark			& $\parallel$ diffusive, good agreement\tablenotemark{b} \cmark	& $\kparr$ too large \xmark \\
						& $\perp$ superdiffusive ($\kperp \to \infty$) \xmark				& $\kperp = \kappa_{\text{FLRW}}$ \xmark	& $\perp$ superdiff. ($\kperp \propto t$) \xmark \\
\hline
\multirow{2}{*}{NLGC\tablenotemark{c}}		& \multirow{2}{4.5cm}{$\kperp$ too high, but scaling with $R$ and $\delta B/B_z$ \cmark}	& ?					& \multirow{2}{*}{$\perp$ diffusive \cmark} \\
						&													&									& \\
\tableline
\multirow{2}{*}{WLNT}		& \multirow{2}{4.5cm}{``Serious mathematical issues''~\citep{2006JPhG...32..809T}}		& $\parallel$ diffusive \cmark				& \multirow{2}{*}{$\perp$ diffusive \cmark} \\
						&													& $\perp$ subdiffusive \cmark				&  \\
\tableline
\end{tabular}
\tablenotetext{a}{Note that there have been hints for subdiffusion at low rigidities~\citep{Casse:2001be,2004JCAP...10..007C}.}
\tablenotetext{b}{Except for steep turbulence spectra where $90^{\circ}$ degree scattering becomes important}
\tablenotetext{c}{NLGC theory requires $\lambda_{\parallel}$ as an input.}
\end{table*}

\subsection{CR anisotropies and backtracking}

Another application of test particle simulations is the study of anisotropies. These are motivated by observations both on large-scale and small-scale anisotropies that hint at limitations of the standard diffusive picture of Sec.~\ref{sec:diffusion_approximation}.

In this standard picture, a small spatial gradient in the CR phase space density leads to the formation of a small dipole in the arrival directions, aligned with the direction of the regular or mean magnetic field. What matters for the formation of the dipole is the gradient over a few mean-free paths before observation and any anisotropy imprinted at larger distances will be destroyed by pitch-angle scattering. However, the phase space density $f$ in the \emph{actual} realisation of the turbulent field will in general differ from the ensemble average $\avg{ f }$, see the discussion in Sec.~\ref{sec:derivation_FPE}, and therefore, also the arrival directions seen by an observer will differ from the dipole predicted for the ensemble-averaged phase space density.

This reasoning has been applied by~\citet{Mertsch:2014cua} to the CR anisotropy problem~\citep{2005JPhG...31R..95H,Zirakashvili:2005gz,Erlykin:2006ri,2006AdSpR..37.1909P,Blasi:2011fm,Evoli:2012ha,Pohl:2012xs,Sveshnikova:2013ui,Kumar:2014dma,Schwadron2014,Ahlers:2016njd}, that is the discrepancy between the measured dipole anisotropy and the one predicted in isotropic diffusion models. Test particle simulations can be used to explore the deviations of the phase space density and anisotropies from the ensemble average in particular realisations of the turbulent magnetic field.

To this end, particles are followed backward in time, starting at position $\vec{r}_{\oplus}$ at time $t$ of observations and computing the trajectories back to an earlier time $t_0$. For a given set of trajectories $\{ \vec{r}_j \}$ from test particle simulations, we can then use Liouville's theorem, that is the conservation of phase space density along trajectories, to connect the phase space density seen by an observer at time $t$ and at the origin of the trajectories $\vec{r}_{\oplus}$ to the assumed phase space density $f(t_0)$ at the other end of the trajectories. More specifically,
\begin{equation}
f(\vec{r}_{\oplus}, \vhat{p}_i(t), t) \simeq f(\vec{r}_i(t_0), \vhat{p}_i(t_0), t_0) \, ,
\label{eqn:backtracking_Liouville}
\end{equation}
where $\vec{r}_i(t')$ and $c \vhat{p}_i(t')$ are the positions and velocities of a particle with position $\vec{r}_i(t) = \vec{r}_{\oplus}$ and velocity $c \vhat{p}_i(t)$ at observation. In order to predict the phase space density seen by an observer at time $t$, some assumptions need to be made on the phase space density at the other ends of the trajectories, specifically at time $t_0$. Usually, for $f(t_0)$ the random fluctuations are ignored and the ensemble-averaged $\avg{ f(t_0) }$ is adopted. Eq.~\eqref{eqn:backtracking_Liouville} then becomes exact if the backtracking time $(t - t_0) \to \infty$. This is motivated by the fact that ensemble averages of second moments of the phase space density, e.g.\ the dipole amplitude or the angular power spectrum, are insensitive to the fluctuations $\delta f$ at $t_0$~\citep{Ahlers:2015dwa}. For the ensemble-average a solution of the CR transport equation is adopted, e.g.\ a spatial gradient.

\begin{figure}[t]
\includegraphics[width=\columnwidth]{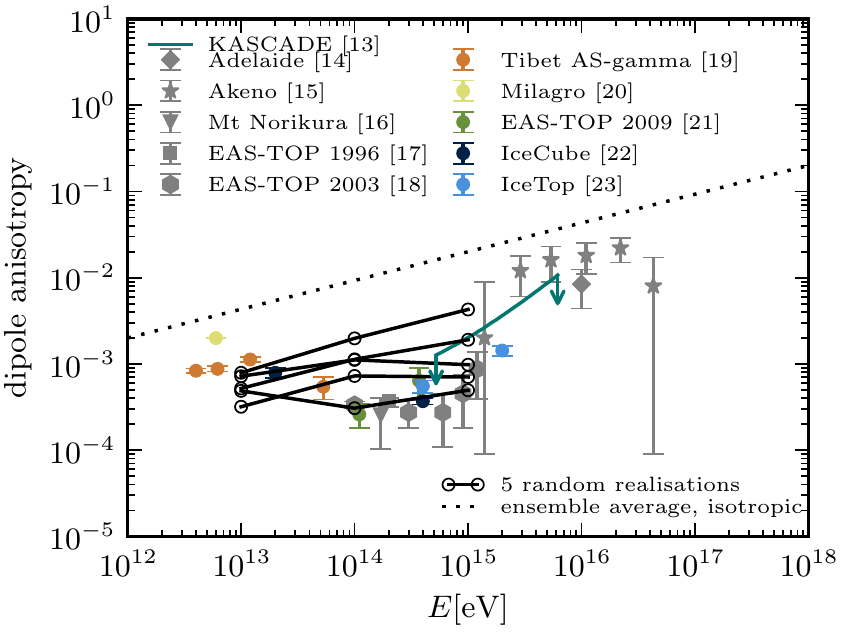}
\caption{The dipole amplitude of Galactic cosmic rays (black open symbols) for five different realisations of the turbulent magnetic field under the assumption of misalignment of background field and cosmic ray gradient. Also shown are some measurements and the expectation from an isotropic diffusion model. From~\citet{Mertsch:2014cua}.}
\label{fig:Mertsch_Funk_2014}
\end{figure}

We show the dipole amplitudes computed with test particle simulations for five different realisations of the turbulent magnetic field in Fig.~\ref{fig:Mertsch_Funk_2014}. It was shown that the intermittency effects due to the turbulent magnetic field can lead to a significant uncertainty in the prediction of the dipole amplitude and direction, both for the case without and with strong background field~\citep{Mertsch:2014cua}. Together with the projection effect due to a potential misalignment between CR gradient and magnetic field direction, this can bring the predicted dipole anisotropy back into agreement with the observations.

The same backtracking technique and Liouville's theorem can be used to also investigate the appearance of anisotropies on small scales~\citep{Abdo:2008kr,Abbasi:2011ai,Abeysekara:2014sna,Aartsen:2016ivj,Abeysekara:2018qho,Aartsen:2018ppz} due to intermittency effects in small-scale turbulence~\citep{Giacinti:2011mz,Ahlers:2015dwa,Lopez-Barquero:2015qpa,Pohl:2015fdp,Kumar:2018qwa}. We refer the interested reader to the recent review by~\citet{Ahlers:2016rox}.

\subsection{The validity of Liouville's theorem}
\label{sec:validity_Liouville}

It has been questioned whether backtracking can be used reliably to investigate the formation of (small-scale) anisotropies~\citep{Lopez-Barquero:2016wnt} and whether Liouville's theorem is valid in the presence of pitch-angle scattering. We therefore provide a few comments on its validity.

First, we note that pitch-angle scattering is to be distinguished from collisions. In collisions the particle trajectories changes abruptly due to short-range forces, e.g.\ hard-sphere collisions in gas kinetic theory. In contrast, in collisonless plasmas each interaction between the particle and a wave-packet changes the particle's pitch-angle only very moderately due to the small turbulent magnetic field, $\delta B^2 / B_z^2 \ll1$ (e.g.~\citealt{2005ppfa.book.....K}). Thus, interactions with many wave-packets are needed for a particle to scatter (which can be defined as a particle changing direction by $180^{\circ}$). The particle trajectories are smooth since the Lorentz force mediating this change is differentiable.

Second, the validity of Liouville's theorem is not only the basis for numerical backtracking, but is also at the heart of kinetic theory, including QLT and its non-linear extensions. If Liouville's theorem was not applicable to collisionless plasmas in the presence of small-scale turbulence, then we would also need to abandon the majority of microscopic particle transport theories and much of plasma theory, in fact.

It has been claimed~\citep{Lopez-Barquero:2016wnt} that conservation of phase space density is equivalent to the conservation of the magnetic moment $M = m v_{\perp}^2 / (2 B)$ of individual particles which can be checked by simulating test particles in random (electro)magnetic fields. We have elsewhere already argued against this view~\citep{Ahlers:2016rox}: While conservation of phase space density requires only differentiability of forces, conservation of the magnetic moment requires the magnetic field to change only adiabatically, that is $B / |\nabla B| \gg r_{\text{g}}$ and $B / \dot{B} \gg \Omega^{-1}$ where $r_{\text{g}}$ and $\Omega$ are the gyroradius and gyrofrequency. Therefore, the conditions for the conservation of the magnetic moment are stricter and variability of the magnetic moment does not imply violation of Liouville's theorem. Note that, of course, magnetic moment $M$ and pitch-angle cosine $\mu$ are closely related for fixed particle energy, such that any pitch-angle scattering necessarily implies the violation of magnetic moment~\citep{2012PhRvE..86a6402D,2015ApJ...811....8W}. The validity of Liouville's theorem is however not affected by this.

Due to the equivalence of phase space volume and (negative) information entropy, it can be said that in the ensemble-average information is lost. The increase of entropy also implies that the evolution of the system is irreversible, reflecting the diffusive nature of the process. However, it is important to realise that the loss of reversibility only occurs through the ensemble averaging. By contrast, in one particular realisation of the turbulent magnetic field, even though particles scatter, phase space volume is conserved, entropy does not increase and the equations of motion are reversible. It is possible to confirm this fact in numerical test particle simulations.

\section{Summary and outlook}
\label{sec:conclusion}

In this \textit{review}, we have given an overview over test particle simulations of CRs that are used to check transport theories, compute their parameters and predict observables beyond the current reach of such theories. In the first part, we summarised the findings of the current paradigm theory, QLT, and its possible extensions. In deriving the Fokker-Planck eq.~\eqref{eqn:Fokker_Planck_simple} and the diffusion eq.~\eqref{eqn:diffusion_equation}, we have reviewed the salient features of QLT, that is the evaluation of the force due to the turbulent magnetic field along unperturbed trajectories and the hierarchy of time scales involved. We have introduced the three most popular analytical turbulence geometries (3D isotropic, slab and composite) and, as an example, have reviewed the derivation of the pitch-angle diffusion coefficient in slab geometry with a broken power law turbulence spectrum. Pointing out some of the shortcomings of QLT, in particular the so-called $90^{\circ}$ problem, we have motivated the need to go beyond the simplest quasi-linear theories. For non-linear theories of CR transport, we have mostly limited ourselves to the BAM model, to NLGC theory and to WLNT.

The second part of this \textit{review} was concerned with test particle simulations itself. First, we developed a technical but central part of running test particle simulations: the generation of the turbulent magnetic field. We have reviewed the two approaches that are regularly used, the harmonic method and the grid method. Both have advantages and disadvantages, but the grid method allows for a much faster evaluation if a large dynamic range in wavenumbers is to be considered. This is particularly true for the nested grid method. We have concluded by reviewing some of the applications of test particle simulations, the major motivation being the current lack of an agreed-upon microscopic transport theory that addresses the various issues that point beyond QLT. Extensions of QLT need to be tested against observations or simulations. Any theory necessarily relies on a certain turbulence model and since the nature of turbulence in the interstellar medium (to a lesser extent also in the interplanetary medium) is uncertain, comparing analytical approaches and numerical simulations based on the same assumed turbulence model is most reliable. We have sketched two important application cases, that is the computation of transport coefficients and the investigation of anisotropies. In doing so, we have stressed the validity of Liouville's theorem for the phase space density before ensemble-averaging which is the basis not only of the backtracking used in anisotropy studies, but also of the analytical approaches.

Over the 25 years since their first use in CR transport studies, test particle simulations have proven a very useful tool.
They have confirmed the sub-diffusive nature of perpendicular transport in slab turbulence, provingthe importance of FLRW. 
Furthermore, parallel transport in isotropic turbulence has been shown to be diffusive where QLT predicts infinite mean-free paths.
Test particle simulations have also allowed for tests of non-linear extensions of QLT, however, with no clear winner yet. 
Despite QLT's deficiencies \emph{in detail}, test particle simulations have reproduced some of its results in a \emph{qualitative} fashion, e.g\ scaling of parallel mean-free path with rigidity and turbulence level, justifying to a certain extent the use of such scalings in phenomenological applications to Galactic CRs. More recently and thanks to increased computing power they have proven helpful in addressing phenomenological issues, for instance the transport at the transition from the resonant to the small-angle scattering regime~\citep{Giacinti:2011ww} or the interpretation of cosmic ray small-scale anisotropies~\citep{Ahlers:2015dwa}.

Open questions that should be further studied and addressed with test particle simulations are the decorrelation of trajectories which leads to the broadening of the resonance condition in non-linear extensions of QLT, the transition in transport from the ballistic to the diffusive regime and a more detailed understanding of CR anisotropies.

On the technological side, a number of improvements are needed to allow for a broader use of test particle simulations though.
It is widely accepted that turbulence is anisotropic in the presence of a background magnetic field. The direction of the anisotropy is determined by the \emph{effective} large-scale field seen at a particular point and on a particular spatial distance scale. Yet, there have been no implementations for the generation of synthetic turbulence with anisotropies resembling those observed, for instance, in MHD simulations. The difficulty here is to allow for the direction of the anisotropy to vary over the spatial domain. The only cases we are aware of~\citep{Giacinti:2017dgt,Demidem:2019jzn} have considered low turbulence levels, such that the direction of the anisotropy is effectively the same at all positions and on all spatial scales.
(Of course, this problem does not arise if the magnetic field is the result of MHD simulations.) 
However, we stress that in the spirit of keeping turbulence physics and transport physics apart, it would be valuable to have such a prescription.
In addition, this would allow to cover a larger dynamical range as with MHD simulations.

Another trend, that is imminent in our view, is the adoption of computing architectures other than CPUs which is what most previous codes have been focussed on. The solution of a large number of equations of motion is perfectly amenable to single instruction, multiple data architectures like graphic processing units (GPUs). The addition of a large number of wave modes needed in the harmonic approach is another example~\citep{2016NewA...45....1T}. 

Given the conceptual simplicity of test particle simulations of CR transport and the availability of computational resources necessary, test particle simulations are thus one of the most important computational tools in studies of CR transport. It is however also necessary to point to the limitations of test particle simulations. First, as alluded to above, the questions of whether the results can be compared to data is hindered by our ignorance of the underlying turbulence model. Of course, analytical transport theories suffer from the same shortcoming. Turning this argument around, we can however hope to constrain the nature of magnetised interstellar turbulence by comparing the results from test particle codes with observations, for example for anisotropies. Also, with ever increasing computational resources, computing trajectories in simulated turbulence will become increasingly important, but for the time being synthetic turbulence is more useful in investigating a number of phenomenological questions.

Second, test particle simulations ignore feedback of the cosmic rays onto the magnetised turbulence, by definition. Approaches like particle-in-cell simulations are appropriate for studying such processes \emph{in principle}, but for the application of such instabilities to astrophysical phenomena, the large dynamical range between plasma skin widths and the relevant astrophysical scales is still challenging. We believe that careful hybrid approaches, combining kinetic cosmic rays with magnetohydrodynamic background plasma will prove most fruitful.

Another open question is the nature of MHD turbulence itself. While the Goldreich-Sridhar picture is an often employed model for anisotropic turbulence, it is based on the assumption of so-called critical balance, meaning that the Alfv\'enic and cascade times are identical. We note that this assumption of a single time-scale is regularly contested in the literature, see e.g.~\citep{2019PhPl...26l2301L}.

\acknowledgments

The author would like to thank the editors of this Topical Collection for the invitation to write this \textit{review} article. The author is also grateful to Andrea Vittino and Marco Kuhlen for reading an earlier version of the manuscript.

\sloppy
\bibliographystyle{spr-mp-nameyear-cnd}  
\bibliography{transport}                

\end{document}